\documentstyle[aps,prl,multicol,epsf]{revtex}
\begin{document}
\draft
\title{Dynamics and Critical Behaviour of the q-model}
\author{Marta Lewandowska$^{1}$, H. Mathur$^{1}$ and Y.-K. Yu$^{2}$} 
\address{$^{1}$Department of Physics, Case Western Reserve University,
Cleveland, OH 44106-7079} 
\address{$^{2}$Department of Physics, Florida Atlantic University,
Boca Raton, FL 33431, USA}


\date{today}
\maketitle
\begin{abstract}
The $q$-model, a random walk model rich in behaviour and applications,
is investigated. We introduce and motivate the $q$-model via its application
proposed by Coppersmith {\em et al.} to the flow of stress through granular
matter at rest. 
For a special value of its parameters the $q$-model has a
critical point that we analyse. To characterise the critical point we imagine
that a uniform load has been applied to the top of the granular medium 
and we study the evolution with depth of fluctuations in the distribution
of load. Close to the critical point explicit calculation 
reveals that the evolution of load exhibits 
scaling behaviour analogous to thermodynamic critical phenomena. The
critical behaviour is remarkably tractable: the harvest of analytic results
includes scaling functions that describe the evolution of the variance
of the load distribution close to the critical point and of the 
entire load distribution right at the critical point, values of the
associated critical exponents, and determination of the upper critical
dimension. These results are of intrinsic interest
as a tractable example of a random critical point. 
Of the many applications of the q-model, the critical behaviour
is particularly relevant to network models of river basins, as
we briefly discuss. Finally we discuss
circumstances under which quantum network models that describe the surface
electronic states of a quantum Hall multilayer can be mapped onto the
classical $q$-model. For mesoscopic multilayers of finite circumference
the mapping fails; instead a mapping to a ferromagnetic supersymmetric
spin chain has proved fruitful. We discuss aspects of the superspin mapping
and give a new elementary derivation of it making use of operator
rather than functional methods.
\end{abstract}
\pacs{PACS: }
\begin{multicols}{2}
\input epsf

\section{Introduction}

It is fortunate that in physics the
same equations sometimes arise in contexts that are 
apparently very different. Feynman illustrates this
through elementary examples in his introductory lectures
on physics to impart the lesson that the ``same equations have the
same solutions'' \cite{feynman}. Our purpose is to study a model,
recently dubbed the $q-$model, that provides another such instance.
The $q-$model has been used to describe the
merging of tributaries to form rivers \cite{scheidegger}; 
the aggregation of diffusing
charges \cite{takayasu}; 
the flow of stress in a granular medium \cite{coppersmith}; and can be mapped
onto the abelian sandpile, a model studied in context of self-organised
criticality \cite{dhar}. It is also closely related to models that describe the
surface of a quantum Hall multilayer \cite{saul,chalker} and passive scalar
turbulence \cite{claudin,siggia}. 
Here we focus on the application
to granular matter, river networks and the quantum Hall multilayer.

Granular matter exhibits fascinating behaviour that
is little understood \cite{reviews}. Examples of granular matter include
sand, powders and agricultural grains stored in silos. An 
important problem is the propagation of
stress through a granular medium at rest. This has been studied by
ingenious experiments, in which a vertical load is applied to an 
amorphous pack of beads, and the loads on the beads in the top
and bottom layers are recorded using carbon paper \cite{science,jaeger}. 
Such experiments
yield the distribution of load on the beads and reveal that there are
no horizontal correlations in load even amongst neighbouring beads.
The $q$-model was introduced by Coppersmith and coworkers to account
for the distribution of load \cite{coppersmith}. 
As we shall see, it also correctly
predicts the lack of horizontal correlation.

For simplicity we describe the $q$-model in a plane.
Since the vertical and horizontal directions are treated
asymmetrically we call this the 1+1 dimensional $q$-model.
The extension to 2+1 dimensions (relevant to experiments on
bead packs) and higher, is straightforward and is discussed 
in section V. In the $q$-model it is assumed that the beads
sit on a regular lattice shown in fig 1. The location of
the beads is specified by the co-ordinates $t$ (the depth of
the layer) and $n$ (the location of the bead within the layer).
Note that $n$ takes only even values for $t$ even; only
odd, for $t$ odd. Each bead is assumed to be supported by
its two nearest neighbours in the layer directly below.
More precisely, it is assumed that a random fraction $f_n(t)$
of the load of bead $(n,t)$ is supported by the neighbour to
the left, bead $(n-1, t+1)$; the remainder, $1 - f_n(t)$,
by the neighbour to the right, bead $(n+1,t+1)$. Denoting
the load on a bead $w$ and its weight $I$ we may write
\begin{eqnarray}
w_n (t)&  = & w_{n-1} (t - 1) [ 1 - f_{n-1} (t-1) ] 
\nonumber \\
& & + w_{n+1} (t-1)
f_{n+1} (t-1) + I_n (t).
\end{eqnarray}
The content of eq (1) is that the load on a bead is the sum
of the loads transmitted to it by its neighbours in the layer above
plus its own weight. The last term in eq (1) is called the 
injection term. Once the fractions are specified, a given
load on the top layer can be propagated downward by use of eq (1).

In the $q$-model it is assumed that the fractions are independent,
identically distributed random variables. The distribution is
assumed to be symmetric about $f=1/2$ to avoid introducing a
horizontal drift to the flow of stress; in other words it is assumed
$P(f) = P(1-f)$. There is no other restriction. Thus the
$q$-models really constitute an enormous family of models
corresponding to different symmetric distributions $P(f)$. To fully
specify a particular model it is necessary to choose the
distribution $P(f)$. One obvious possibility is to take
$P(f)$ to be uniformly distributed over the unit interval;
another is to assume that the fractions must be 0 or 1 with
equal probability. The latter is called the singular distribution.

Mathematically, the $q$-model is a problem of random walkers that
coalesce upon contact and fission spontaneously. The singular distribution
corresponds to the case that the walkers coalesce but do not fission.

Coppersmith {\em et al.} argued that, neglecting injection,
at sufficient depth the distribution
of load would attain a steady state \cite{coppersmith}. 
They studied $\Pi (w, t \rightarrow
\infty)$, the probability distribution of load on beads in a sufficiently
deep layer. For almost 
all distributions $P(f)$, except the singular distribution,
they concluded that $\Pi( w, t \rightarrow \infty)$ decays exponentially
for large $w$. This agrees with experiment and constitutes an important
success of the $q$-model. For the singular distribution, Coppersmith
{\em et al.} argued that $\Pi(w,\infty)$ follows a power law. Hence
they conjectured that the singular distribution constitutes a critical
point in the family of $q$-models. A major goal of this paper is to make
this analogy to thermodynamic critical phenomena precise by detailed
analysis of the critical point.

In spite of the success mentioned above the $q$-model cannot be
considered a complete theory of stress propagation in granular matter.
This is clear both empirically and on grounds of internal consistency.
Since the publication of the $q$-model, interesting new ideas on the
subject of stress flow have appeared \cite{claudin,narayan,liu,socolar}, 
but in this paper we restrict 
attention to the
$q$-model. This seems justified because the $q$-model does capture
some elements of the physics correctly and because it exhibits non-trivial
critical behaviour that is interesting in its own right.

Further motivation to study the $q$-model and particularly its
critical point comes from hydrology. To make contact with that 
subject consider a singular $q$-model with zero injection
and imagine that only a few
beads in the top layer are loaded. The load then zig-zags downwards,
perhaps along the lines shown in fig 2. If we interpret these lines as
tributaries merging to form a river we arrive at Scheidegger's 
model \cite{scheidegger}
which appeared in the hydrology literature more than thirty
years ago \footnote{Parenthetically we note that Scheidegger's
model is purely descriptive in the sense that it is a recipe to 
draw statistically realistic networks. Somewhat different in spirit
are models that seek to represent physical processes, sometimes
very crudely, by which the network forms. Two examples of
such models in the recent Physics literature are refs \cite{mardar,leheny}.
The model of Leheny and Nagel for example describes an apocalyptic 
lattice world with discrete time. Each time step brings precipitation,
and in its wake, erosion and avalanches. Realistic networks result.}. 
Networks of tributaries in 
river basins are known empirically to be scale invariant structures that
obey a variety of
power laws. Scheidegger networks too obey these laws and are in this
statistical sense extremely realistic representations of 
river basins. An excellent discussion of river
basin power laws is given in refs \cite{riverbook,riverpower}. 
Ref \cite{riverbook}
presents some discussion of data; ref \cite{riverpower} provides a detailed
comparison between real and Scheidegger networks. 

Here we wish to point out that non-singular $q$-models too can be
interpreted as models of river networks. For example, consider a model
in which the fractions can take only the values 0, 1/2 and 1 with 
probability $(1-\delta)/2$, $\delta$ and $(1-\delta)/2$ respectively.
This model reduces to Scheidegger's as $\delta \rightarrow 0$. It produces
networks similar to Scheidegger's except that occasionally streams split
to form distributaries. Thus this network is topologically distinct
from Scheidegger networks. More significantly, as we show below, a network
with non-zero $\delta$ is not scale-invariant. This is reminiscent of
a river network model studied by Narayan and Fisher \cite{onuttom}. 
In their ``rocky-river''
model too the network is not scale invariant except if a model parameter
is tuned to a special (critical) value. Effectively this tuning parameter
also controls river splitting. Taken together, these results suggest
that river splitting is a relevant perturbation that spoils the scale
invariant structure of networks. In this paper we concentrate on showing
that $q$-model networks with river splitting are not scale invariant. We do not
explore whether such non-scale invariant networks are realised in nature
(for further discussion and speculation in this regard, however, 
see section VII). 

A quantum Hall multilayer consists of layers of two-dimensional
electron gases stacked vertically. Multilayers can be realised by 
fabricating an appropriate GaAs heterostructure \cite{stormer}. They are also
realised naturally in some organic salts. In a quantum Hall multilayer a 
sufficiently large magnetic
field is applied perpendicular to the layers so that the lowest
Landau level in each layer is fully occupied. Under this circumstance
the only important electronic states in each layer are the chiral
edge states that propagate in one direction only as shown in Fig 3(a).
These edge states are coupled by tunneling between layers. Thus the
surface of a multilayer is covered by a chiral sheath of coupled
edge states. These surface states control the electrical transport
properties of the multilayer. A central question from a quantum 
transport point of view is whether these surface states are localized
or extended in the direction of the field \cite{chalker,matthew,ilya}.

Fig 3(b) shows a network model of the multilayer surface introduced
by Saul, Kardar and Read \cite{saul} and studied by many authors subsequently.
In this model it is assumed that tunneling between edges takes place
only at discrete nodes (dashed vertical lines in Fig 3b) that appear
at regular intervals along an edge. The edges are separated by nodes 
into horizontal segments called links. The wavefunction has a definite 
value on each link. Each node is visited by two incoming links and by
two outgoing links. Each node is characterised by a $2 \times 2$ S-matrix
that relates the wavefunction on the outgoing links to the incoming 
amplitudes. Once the S-matrices are specified, given the wavefunction
through a vertical slice, we can propagate it to the right. The S-matrices
are chosen at random from some ensemble to incorporate the effect of
disorder. To fully specify the model it is necessary to choose a 
distribution for the S-matrices. Periodic boundary conditions are 
imposed in the horizontal direction \cite{chalker}.

The directed network model above is quantum mechanical but in
the limit of infinite circumference and for a special choice of
disorder, Saul, Kardar and Read have shown that it reduces to a 
{\em classical} model, the $q$-model with uniform distribution of
fractions and zero injection \cite{saul}. 
In section VI we discuss some respects
in which more generic models of the multilayer surface, that do not
reduce to classical models, still do show behaviour similar to the
$q$-model \cite{heisenberg,multifractal}. 
At the same time we show that in case of finite circumference
quantum interference effects become important and there is little
to be learnt from the study of the classical $q$-model. Instead a mapping to a
ferromagnetic supersymmetric spin-chain has proved fruitful in this
case \cite{ilya,zirnbauer}. In section VI we discuss aspects of this
mapping.

A detailed summary of our results is given in section VII. The reader
interested in first obtaining an overview of the paper or interested
only in the results should proceed directly to section VII.

\section{Critical Behaviour in 1+1 Dimensions}

Coppersmith {\em et al.} analysed the distribution of load
$\Pi (w,t \rightarrow \infty)$ at very large depth where presumably a 
steady state is achieved \cite{coppersmith}. 
Here we study how the distribution evolves as 
a function of depth to this asymptotic steady state. We assume that
a uniform load is applied to the top layer,
\begin{equation}
w_n (t) = 1 \hspace{4mm} {\rm for} \hspace{2mm} {\rm all}
\hspace{2mm} n.
\end{equation}
In this section we neglect the weight of each bead (the injection
term). In partial support of this neglect  we note that in the experiment
of ref \cite{jaeger} typically 
a total load of 7600 N was applied to the bead pack. In
comparison we estimate that the weight of a single bead was
less than a mN; of the entire pack, less than 100 N. 
However, right at the critical point injection is a relevant
perturbation, and at sufficiently large depth must be taken into account
even if the weight of a single bead is small. We return to the effects of
injection in section IV.

To make the problem tractable we study not the entire distribution
$\Pi(w,t)$ but only its lowest non-trivial moment. With the neglect
of injection it follows that the total load on every layer is the
same; the $q$-model dynamics (eq 1) just shuffles this load. Hence
the average load in layer $t$
\begin{eqnarray}
\langle w(t) \rangle & = & \int_{0}^{\infty} d w w \Pi(w,t) 
\nonumber \\
 & = & 1.
\end{eqnarray}
The lowest non-trivial moment is therefore the variance
\begin{equation}
\langle \delta w^2 (t) \rangle = \int_{0}^{\infty} d w w^2
\Pi(w,t) - 1.
\end{equation}
Since a uniform load is applied to the top layer the variance
in that layer vanishes. As the load propagates downward, the
fluctuations must grow and saturate. Our purpose is to analyse
this evolution for different distributions $P(f)$, particularly
those that are close to the singular distribution.

Right at the critical point the asymptotic distribution $\Pi(w,\infty)$
is believed to be a power law. If we assume that it does not have a
well defined variance, then by analogy to critical phenomena we surmise
that close to the critical point the variance
must diverge as
\begin{equation}
\langle \delta w^2 (t \rightarrow \infty) \rangle \sim
\frac{1}{\delta^{\theta}}.
\end{equation}
Here $\delta$ measures the distance of a distribution $P(f)$ from
the singular distribution; $\delta$ will be defined precisely below.
We also expect that the depth-scale $\xi_{{\rm corr}}$ 
at which the steady state
is attained will diverge as the critical point is approached. Thus
\begin{equation}
\xi_{{\rm corr}} \sim \frac{1}{\delta^{\varphi}}.
\end{equation}
$\xi_{{\rm corr}}$ is a vertical correlation length that diverges as 
the critical
point is approached. Combining eqs (5) and (6) we expect that close
to the critical point the fluctuations must have a scaling form
\begin{equation}
\langle \delta w^2 (t) \rangle = \frac{1}{\delta^{\theta}}
{\cal F} ( t \delta^{\varphi} ).
\end{equation}
To be consistent with eq (5) we expect that the scaling function
${\cal F} (u) \rightarrow $ const as $ u \rightarrow \infty$. For short times
we expect that the system should behave as it would at the critical point. 
The $\delta$ dependence should cancel and so we expect 
${\cal F} (u) \sim u^{\theta/
\varphi}$ for $ u \ll 1$ so that $\langle \delta w^2 (t) \rangle \sim 
t^{\theta/\varphi}$ at the critical point.

In the remainder of this section we will confirm that eq (7) and these
inferences are valid. We will determine the exponents $\theta$ 
and $\varphi$ and the scaling function ${\cal F} (u)$.

As an aside to experts we note that it may have been more natural
to name the exponents $\theta \rightarrow (3 - \tau)/\sigma$ and
$\varphi \rightarrow \nu z$. These names follow from a more general
scaling hypothesis for the entire distribution (eq 174). However 
in this section we have elected to make the more restricted hypothesis
eq (7) and to give the exponents single letter names taking care
to avoid common exponent names such as $\alpha, \beta$ and $\nu$.

\subsection{Disorder Average}

Consider the correlation function
\begin{equation}
c_m (t) = \frac{1}{N} \sum_{n} \langle w_n (t) w_{n+m} (t) \rangle.
\end{equation}
We assume there are $N$ beads in each layer and we impose periodic 
boundary conditions in the horizontal direction. Ultimately
we are interested in taking $N \rightarrow \infty$. Note that $m$
is even for both $t=$ even and $t=$ odd. In terms of the correlation
function the variance is given by
\begin{equation}
\langle \delta w^2 (t) \rangle = c_{0} (t) - 1.
\end{equation}
The correlation function obeys a remarkably simple evolution equation.
This equation can be solved by straightforward classical analysis to
yield the evolution of the variance. It is not difficult to obtain the
entire correlation function by this method, and thereby obtain information
on the horizontal correlation length, but we do not attempt this here.

To analyse the evolution of the correlation function we write
\begin{eqnarray}
c_{m} (t+1) & = & \frac{1}{N} \sum_{n} \langle w_{n} (t+1) w_{n+m} (t+1)
\rangle \nonumber \\
 & = & \frac{1}{N} \sum_{n} \langle w_{n+1} (t) w_{n+m+1} (t) 
f_{n+1}(t) f_{n+m+1} (t) \rangle \nonumber \\
 & & + \hspace{2mm} {\rm others}. \nonumber \\
 & = & \frac{1}{N} \sum_{n} \langle w_{n+1} (t) w_{n+m+1} (t) \rangle
\langle f_{n+1}(t) f_{n+m+1} (t) \rangle \nonumber \\
 & & + \hspace{2mm} {\rm others}.
 \end{eqnarray}
To obtain the second line of eq (10) we have used eq (1). Four terms
result; we have written only one for illustration. 
To obtain the third line it is crucial to observe
that $w_{n}(t)$ depends only on fractions in the layers
above. It is not correlated with the fractions in layer $t$,
allowing us to factorise the average as shown.

To perform the average we need information about the distribution
$P(f)$. By symmetry for any choice of distribution
\begin{equation}
\langle f \rangle = \int_{0}^{1} df f P(f) = \frac{1}{2}.
\end{equation}
For the variance we write
\begin{equation}
\langle \left( f - \frac{1}{2} \right)^2 \rangle = \frac{\epsilon}{4}.
\end{equation}
$\epsilon$ is a parameter that characterises the distribution
$P(f)$. For example, $\epsilon = 1/3$ for the uniform distribution. 
For the singular distribution the parameter takes its maximum possible
value $\epsilon = 1$. Since the fractions
for different beads are assumed to be independently distributed we
conclude
\begin{equation}
\langle f_n (t_1) f_m (t_2) \rangle = \frac{1}{4} + 
\frac{\epsilon}{4} \delta_{n,m} \delta_{t_1, t_2}.
\end{equation}

Substituting eq (13) in eq (10) we obtain
\begin{eqnarray}
c_m (t+1) & = & \left( \frac{1}{4} + \frac{\epsilon}{4} \delta_{m,0}
\right) c_m (t) \nonumber \\
 & & + \hspace{2mm} {\rm others} \nonumber \\
 & = & \left( \frac{1}{2} + \frac{\epsilon}{2} \delta_{m,0} \right) c_m (t)
\nonumber \\
& & + \left( \frac{1}{4} - \frac{\epsilon}{4} \delta_{m,2} \right) 
c_{m-2} (t)
\nonumber \\
& & + \left( \frac{1}{4} - \frac{\epsilon}{4} \delta_{m,-2} \right) 
c_{m+2} (t).
\end{eqnarray}
In the second line of eq (14) the other terms have been unveiled.
Recall that $m$ takes even integer values. It is convenient to
replace $ m \rightarrow m/2$ to obtain
\begin{eqnarray}
c_m(t+1) & = & \left( \frac{1}{2} + \frac{\epsilon}{2} \delta_{m,0} \right) 
c_m (t)
\nonumber \\
 & & + \left( \frac{1}{4} - \frac{\epsilon}{4} \delta_{m,1} \right) 
c_{m-1} (t) 
\nonumber \\
& & + \left( \frac{1}{4} - \frac{\epsilon}{4} \delta_{m,-1} \right) 
c_{m+1} (t).
\end{eqnarray}

Eq (15) is the main result of this subsection. It governs the evolution
of the correlation function. We wish to solve it subject to the initial
condition
\begin{equation}
c_m (t \rightarrow 0) = 1 \hspace{2mm} {\rm for} \hspace{2mm} {\rm all}
\hspace{2mm} m.
\end{equation}
The initial condition follows from the definition of $c_m$ (eq 8) and
the assumed uniform load on the top layer. Note that the distribution
$P(f)$ enters the evolution equation only through the parameter $\epsilon$.
Since the parameter takes its maximum value
$\epsilon = 1$ for the singular distribution we may define
\begin{equation}
\delta = 1 - \epsilon
\end{equation}
as the distance of a distribution $P(f)$ from the critical point. 

\subsection{Scattering solution}

It is easy to verify that a steady state solution to eq (15) is
\begin{eqnarray}
c_0 & = & \frac{1}{1 - \epsilon};
\nonumber \\
c_n & = & 1 \hspace{2mm} {\rm for} \hspace{2mm} n \neq 0.
\end{eqnarray}
Assuming this is the unique steady state towards which our initial
condition evolves, eq (18) reveals that the variance does diverge
as the singular distribution is approached. Using eq (9)
\begin{eqnarray}
\langle \delta^2 w( t \rightarrow \infty ) \rangle & = & 
\frac{\epsilon}{1 - \epsilon} 
\nonumber \\
 & \approx & \frac{1}{\delta} \hspace{2mm} {\rm as} \hspace{2mm}
\epsilon \rightarrow 1.
\end{eqnarray}
Comparing eq (5) we see that the exponent $\theta = 1$. Eq (18)
also reveals that in steady state the fluctuations in load are 
uncorrelated for all pairs of beads including neighbours. This is
in agreement with experiment \cite{jaeger}.

A full solution of evolution dynamics needs more work. Schematically 
eq (15) states
\begin{equation}
c (t+1) = H c(t).
\end{equation}
The strategy we adopt here is to seek the eigenvectors of $H$, 
\begin{equation}
H \phi^{\lambda} = \lambda \phi^{\lambda},
\end{equation}
and to expand the initial correlation vector $c(0)$ in terms of
the eigenvectors, 
\begin{equation}
c(0) = \sum_{\lambda} a_{\lambda} \phi^{\lambda}.
\end{equation}
The correlation vector at depth $t$ is then
\begin{equation}
c(t) = \sum_{\lambda} \lambda^t a_{\lambda} \phi^{\lambda}.
\end{equation}

A complication we must negotiate is that $H$ is non-Hermitian.
According to the standard theory of biorthogonal expansion
(briefly recounted in Appendix A) to execute the plan above
we must {\em prove} that the eigenvectors of $H$ span the 
vector space. Then we must find the eigenvectors of $H^{\dagger}$,
called the left eigenvectors of $H$ in this context. The eigenvalues of
$H^{\dagger}$ are the complex conjugate of the eigenvalues of $H$.
Thus
\begin{equation}
H^{\dagger} \psi^{\lambda} = \lambda^{*} \psi^{\lambda}.
\end{equation}
$\psi^{\lambda}$ denotes the left eigenvector with eigenvalue
$\lambda^{*}$. Having completed these tasks we may write the
completeness relation
\begin{equation}
\sum_{\lambda} ( \psi^{\lambda}_{m} )^{*} \phi_{n}^{\lambda} 
= \delta_{mn}.
\end{equation}
Using eq (25) we conclude that the expansion coefficients in
eq (22) are determined by the left eigenvectors:
\begin{equation}
a_{\lambda} = \sum_{m} ( \psi^{\lambda}_{m} )^{*} c_m (0).
\end{equation}

Implementing the plan we first write the eigenvalue equation for
$H$
\begin{eqnarray}
\frac{1}{2} \phi^{\lambda}_r + \frac{1}{4} \phi^{\lambda}_{r+1}
+ \frac{1}{4} \phi^{\lambda}_{r-1} & = & \lambda \phi^{\lambda}_{r}
\hspace{2mm} {\rm for} \hspace{2mm} |r| \ge 2;
\nonumber \\
\frac{1}{2} \phi^{\lambda}_{-1} + \frac{1}{4} \phi^{\lambda}_{-2}
+ \frac{1 - \epsilon}{4} \phi^{\lambda}_{0} & = & \lambda
\phi^{\lambda}_{-1};
\nonumber \\
\frac{1+ \epsilon}{2} \phi^{\lambda}_{0} +
\frac{1}{4} \phi^{\lambda}_{-1} + \frac{1}{4} \phi^{\lambda}_{1}
& = & \lambda \phi^{\lambda}_{0};
\nonumber \\
\frac{1}{2} \phi^{\lambda}_1 + \frac{1 - \epsilon}{4} \phi^{\lambda}_0
+ \frac{1}{4} \phi^{\lambda}_2 & = & \lambda
\phi^{\lambda}_1.
\end{eqnarray}
Note that for $\epsilon = 0$ eq (27) may be interpreted as the
Schr\"{o}dinger equation for a free particle on a tightbinding
lattice, familiar from elementary solid state physics. For non-zero
$\epsilon$ the particle may be viewed as scattering 
off a (non-Hermitian) barrier at 
the origin. Thus we seek a solution of the scattering form
\begin{eqnarray}
\phi^{(+)k}_{n} & = & T(k) e^{i k n} \hspace{2mm} {\rm for} \hspace{2mm}
n \geq 1;
\nonumber \\
 & = & A(k) \hspace{2mm} {\rm for} \hspace{2mm} n=0;
\nonumber \\
 & = & e^{ikn} + R(k) e^{-ikn} \hspace{2mm} {\rm for} \hspace{2mm} n \leq -1.
\end{eqnarray}
Here $0 < k < \pi$. The first line of eq (27) then yields the eigenvalue
\begin{equation}
\lambda (k) = \frac{1}{2} + \frac{1}{2} \cos k.
\end{equation}
The next three lines yield the scattering coefficients
\begin{eqnarray}
A(k) & = & \frac{ i \sin k}{ (1 - \epsilon) e^{ik} + \epsilon - \cos k};
\nonumber \\
T(k) & = & (1 - \epsilon) A(k);
\nonumber \\
R(k) & = & (1 - \epsilon) A(k) - 1.
\end{eqnarray}
There are also scattering solutions to eq (27) corresponding to the
fictitious particle coming in from the right
\begin{eqnarray}
\phi_{n}^{(-)k} & = & e^{-ikn} + R(k) e^{ikn} \hspace{2mm}
{\rm for} \hspace{2mm} n \geq 1;
\nonumber \\
 & = & A(k) \hspace{2mm} {\rm for} \hspace{2mm} n=0;
\nonumber \\
 & = & T(k) e^{-ikn} \hspace{2mm} {\rm for} \hspace{2mm} n \leq -1.
\end{eqnarray}
By symmetry the scattering coefficients for this state are also
given by eq (30).

There are no bound state solutions to eq (27). The scattering solutions
we have found all have real eigenvalues. In principle, since $H$
is non-Hermitian, complex eigenvalues
are also possible. However it turns out there are no solutions with
complex eigenvalue that are biorthonormalisable. It will be seen that
the scattering solutions we have found constitute a complete set.

The next step is to find the left eigenvectors that obey
\begin{eqnarray}
\frac{1}{2} \psi^{\lambda}_r + \frac{1}{4} \psi^{\lambda}_{r+1}
+ \frac{1}{4} \psi^{\lambda}_{r-1} & = & \lambda \psi^{\lambda}_{r}
\hspace{2mm} {\rm for} \hspace{2mm} |r| \ge 2;
\nonumber \\
\frac{1}{2} \psi^{\lambda}_{-1} + \frac{1}{4} \psi^{\lambda}_{-2}
+ \frac{1}{4} \psi^{\lambda}_{0} & = & \lambda
\psi^{\lambda}_{-1};
\nonumber \\
\frac{1+ \epsilon}{2} \psi^{\lambda}_{0} +
\frac{1 - \epsilon}{4} \psi^{\lambda}_{-1} + 
\frac{1 - \epsilon}{4} \psi^{\lambda}_{1}
& = & \lambda \psi^{\lambda}_{0};
\nonumber \\
\frac{1}{2} \psi^{\lambda}_1 + \frac{1}{4} \psi^{\lambda}_0
+ \frac{1}{4} \psi^{\lambda}_2 & = & \lambda
\psi^{\lambda}_1.
\end{eqnarray}
Eq (32) is the transpose of eq (27). The left eigenvectors are
\begin{eqnarray}
\psi^{(+)k}_{n} & = & {\cal T} (k) e^{i k n} \hspace{2mm} {\rm for} \hspace{2mm}
n \geq 1;
\nonumber \\
 & = & {\cal A} (k) \hspace{2mm} {\rm for} \hspace{2mm} n=0;
\nonumber \\
 & = & e^{ikn} + {\cal R} 
(k) e^{-ikn} \hspace{2mm} {\rm for} \hspace{2mm} n \leq -1.
\end{eqnarray}
and
\begin{eqnarray}
\psi_{n}^{(-)k} & = & e^{-ikn} + {\cal R} (k) e^{ikn} \hspace{2mm}
{\rm for} \hspace{2mm} n \geq 1;
\nonumber \\
 & = & {\cal A} (k) \hspace{2mm} {\rm for} \hspace{2mm} n=0;
\nonumber \\
 & = & {\cal T} (k) e^{-ikn} \hspace{2mm} n \leq -1.
\end{eqnarray}
The scattering coefficients are given by
\begin{eqnarray}
{\cal A} (k) & = & \frac{ i (1 - \epsilon) \sin k}{ (\epsilon - 
\cos k) + (1 - \epsilon) e^{ik} };
\nonumber \\
{\cal T} (k) & = & {\cal A} (k);
\nonumber \\
{\cal R}(k) & = & {\cal A} (k) - 1.
\end{eqnarray}

Having found the left and right eigenvectors, by analogy with
eq (25), we now posit the
completeness relation
\begin{equation}
\int_{0}^{\pi} \frac{dk}{2 \pi} \left( 
\psi^{(+)k*}_m \phi^{(+)k}_n 
+ \psi^{(-)k*}_m \phi^{(-)k}_n \right) 
= \delta_{mn}.
\end{equation}
The proof of this completeness relation, an important
element of the analysis, is carried out in Appendix A.

The expansion of the initial correlation vector indicated
schematically in eq (22) may now be written
\begin{equation}
c_m (0) = \int_{0}^{\pi} \frac{d k}{2 \pi} \left[
a^{(+)}(k) \phi^{(+)k}_{m} 
+ a^{(-)} (k) \phi^{(-)k}_{m} \right].
\end{equation}
The correlation vector at depth $t$ is now
\begin{equation}
c_m (t) = \int_{0}^{\pi} \frac{d k}{2 \pi} \lambda(k)^t
\left[a^{(+)}(k) \phi^{(+)k}_{m} 
+ a^{(-)} (k) \phi^{(-)k}_{m} \right]
\end{equation}
as previously shown schematically in eq (23).

The expansion coefficients $a(k)$, obtained using the
completeness relation (eq 36), are
\begin{eqnarray}
a^{(+)}(k)  = \sum_{n=-\infty}^{+\infty} c_n (0) \psi^{(+)k*}_n;
\nonumber \\
a^{(-)}(k)  = \sum_{n=-\infty}^{+\infty} c_n (0) \psi^{(-)k*}_n;
\end{eqnarray}
as previously indicated schematically in eq (26). To ensure convergence of the
sums in eq (39) we set $ c_m (0) \rightarrow e^{- \eta |m|}$ and
take $\eta \rightarrow 0$ at the 
end. Using eqs (33), (34) and (35) we perform the sums exactly to
obtain
\begin{eqnarray}
a^{(+)}(k) & = & a^{(-)}(k) 
\nonumber \\
 & = & {\cal A}(k)^{*} + 2 {\cal A}(k)^{*} 
\frac{ e^{-i k - \eta} }{1 - e^{ - i k - \eta } } +
\left( \frac{ e^{ i k - \eta } }{1 - e^{i k - \eta} } - {\rm cc} 
\right) \nonumber \\
 & = & 2 \pi {\cal A}(k)^{*} \delta (k) + [ 1 - {\cal A}(k)^{*} ]i 
\cot \frac{k}{2}.
\end{eqnarray}
The last line of eq (40) is obtained by taking the limit $ \eta \rightarrow
0$. 

Substituting eq (40) in eq (38) and making use of eqs (28), (29), (30), (31)
and (35) we finally obtain
\begin{equation}
c_0 (t) = \frac{1}{1 - \epsilon} - \frac{\epsilon}{\pi} \int_{0}^{\pi}
d k \frac{ \cos^{2(t+1)} (k/2) }{\epsilon^2 - (2 \epsilon - 1) \cos^2 k}.
\end{equation}
Eq (41) is the exact expression for the evolution of $c_0 (t)$ that we
sought. 

Finally we would like to re-express eq (41) in terms of standard
special functions. Some of the manipulations will prove useful later
in the analysis of injection. Introduce the $z$-transform
\begin{eqnarray}
c_{0} (z) & = & \sum_{t=0}^{\infty} z^t c_{0} (t) 
 = \frac{1}{1 - \epsilon} \frac{1}{1 - z} 
\nonumber \\
 & &  - \frac{\epsilon}{\pi} \int_{0}^{\pi} d k
\frac{ \cos^2 (k/2) }{ \epsilon^2 - (2 \epsilon - 1) \cos^2 (k/2) }
\frac{1}{1 - z \cos^2 (k/2) }.
\end{eqnarray}
The $k$-integral may now be performed\footnote{Extend the range from
0 to $2 \pi$. Then use the standard trick for turning an angular integral
into a contour integral around the unit circle in the $\zeta$ plane via
the substitution $\zeta \rightarrow e^{i k}$. See for example \cite{morse},
p 409.} to yield
\begin{eqnarray}
c_{0}(z) & = & \frac{1}{(1-\epsilon)(1-z)} 
- \frac{\epsilon^2}{(1 - \epsilon)(2 \epsilon - 1) (1 - \alpha(\epsilon) z)}
\nonumber \\
 & &
- \frac{\epsilon}{(1 - 2 \epsilon) \sqrt{1 - z} ( 1 - \alpha( \epsilon ) z )}.
\end{eqnarray}
For brevity $\alpha (\epsilon) = \epsilon^2/(2 \epsilon - 1)$. 
Upon inversion of the $z$-transform (details relegated to Appendix B) we 
obtain
\begin{equation}
c_0 (t) = \frac{1}{1 - \epsilon} - \frac{1}{\pi \epsilon}
\int_{1}^{\infty} d x (x - 1)^{-1/2} \frac{1}{x^{t+1}}
\left(x - \frac{2 \epsilon - 1}{\epsilon^2} \right)^{-1}.
\end{equation}
Comparing an integral representation for the hypergeometric function
\cite{morse}
\begin{equation}
F(a,b,c;s) = \frac{ \Gamma (c) }{ \Gamma(c - b) \Gamma (b) }
\int_{1}^{\infty} d x (x - 1)^{c - b - 1} x^{a - c} (x - s)^{-a},
\end{equation}
valid for Re $c > $ Re $b > 0$ and $|s| < 1$, we conclude
\begin{eqnarray}
\langle \delta w^{2} (t) \rangle& = &
\frac{\epsilon}{1 - \epsilon} 
- \frac{1}{\pi \epsilon} 
\frac{ \Gamma (1/2) \Gamma( t + 3/2 )}{\Gamma (t + 2) }
\nonumber \\
& & \times
F \left(1, t + 3/2, t + 2; \frac{2 \epsilon - 1}{\epsilon^2} \right)
\end{eqnarray}
Eq (46) is the final result of this section. It is an exact formula
for the evolution of fluctuations with depth, in terms of known 
special functions. As a practical matter eqs (41) and (44) are equivalent
to eq (46) and will prove more useful.

\subsection{Scaling Limit}

Eq (46) gives the exact evolution of load fluctuations for the
$q$-model without injection. It is valid for all $t$ and all
distributions, $P(f)$.
From our point of view however it is more interesting to examine 
the scaling limit of large depth behaviour near the critical point.

To derive the scaling limit we start with eq (44)---eq (41) would
have served just as well---and consider the limit $t \gg 1$ and 
$\delta = 1 - \epsilon \rightarrow 0$. We do not make any assumption
about the relationship between $t$ and $1/\delta$. We obtain
\begin{eqnarray}
\langle \delta^2 w (t) \rangle & \approx & \frac{1}{\delta}
- \frac{1}{\pi} \int_{0}^{\infty}d s s^{-1/2} e^{ - t \ln (1 + s) }
(s + \delta^2)^{-1}
\nonumber \\
 & \approx & \frac{1}{\delta} - \frac{1}{\pi} \int_{0}^{\infty}d s s^{-1/2} 
e^{ - t s } (s + \delta^2)^{-1}.
\end{eqnarray}
In the first line of eq (47) we have changed the integration variable
from $x$ to $s = x-1$. Again changing the integration variable from
$s$ to $\overline{s} = s/\delta^2$ we obtain
\begin{equation}
\langle \delta^2 w (t) \rangle = \frac{1}{\delta} \left[ 1 - \frac{2}{\pi}
\int_{0}^{\infty} d \overline{s} \frac{ e^{-\overline{s}^2 t
\delta^2} }{1 + \overline{s}^2 } \right].
\end{equation}
Comparing eq (7) we conclude that close to the critical point and in
the large depth limit $\langle \delta^2 w ( t ) \rangle $ does indeed
have a scaling form with exponents 
\begin{equation}
\theta = 1, \hspace{3mm} \varphi = 2
\end{equation}
and scaling function
\begin{equation}
{\cal F} (u) = 1 - \frac{2}{\pi} \int_{0}^{\infty}
d s \frac{ e^{- u s^2} }{1 + s^2}. 
\end{equation}
Fig 4 shows a plot of ${\cal F} (u)$. As anticipated the asymptotic
behaviour of the scaling function is
\begin{eqnarray}
{\cal F}(u) & \approx & 1 - \frac{1}{\sqrt{ \pi u }} \hspace{2mm}
{\rm for} \hspace{2mm} u \rightarrow \infty
\nonumber \\
 & \approx & \frac{2}{\sqrt{\pi}} \sqrt{u} \hspace{2mm} {\rm for}
\hspace{2mm} u \rightarrow 0.
\end{eqnarray}
We conclude that the saturation depth scale $\xi_{{\rm corr}} \sim 1/\delta^2$. 
For very great depths $t \gg \xi_{{\rm corr}}$, the fluctuations saturate to the
value $1/\delta$ as found earlier by analysis of the steady state
(eq 19). For small depths, $1 \ll t \ll \xi_{{\rm corr}}$ they grow
as 
\begin{equation}
\langle \delta^2 w (t) \rangle = \frac{2}{\sqrt{\pi}} \sqrt{t}.
\end{equation}
This behaviour must persist at all depths right at the critical
point as will be explicitly confirmed in section III.

In summary, we have shown that the singular distribution is an 
isolated critical point in the space of $q$-models. There is a
(vertical) correlation length that diverges as the critical point
is approached. We have determined the
exponents $\theta$ and $\varphi$ and the scaling function ${\cal F}(x)$
introduced in eqs (5), (6) and (7).
In context of river networks we have found that any $q$-model
with stream splitting (hence non-zero $\delta$) has a 
(possibly very long) correlation length in the direction of flow.
Such a network is therefore not scale invariant on sufficiently long
scales\footnote{Strictly, to analyse a river network the appropriate
initial condition is to load a fraction of randomly chosen sites in
the $t=0$ layer, rather than the uniform load analysed here. However
we do not expect our conclusion regarding correlation lengths is
sensitive to initial conditions.}. 

\section{Critical Point Distribution}

Right at the critical point in $1+1$ dimensions it is possible
to analyse the dynamics of the entire distribution $\Pi (w,t)$.
Since there is no vertical length scale at the critical point
we expect that in the large depth, scaling limit
\begin{equation}
\Pi (w,t) = \frac{1}{t^{\omega}} {\cal H} (w t^{\Upsilon}).
\end{equation}
Eq (53) implies that at the critical point the variance should
grow as $t^{-3 \Upsilon - \omega}$ in the scaling limit $ t \gg 1$.
From eq (7) we had surmised that the variance would grow as
$t^{\theta/\varphi}$ for $\delta = 0$. Hence the exponents $\theta,
\varphi$ of the preceding section and $\omega,\Upsilon$ of this section are not
independent; they satisfy the relation $3 \Upsilon + \omega + \theta/\varphi
=0$.  
Below we calculate the exponents $\omega$ and $\Upsilon$, explicitly 
verify the exponent relationship and obtain the scaling
function ${\cal H}(s)$.

Again as an aside to experts we note that the exponents $\omega$
and $\Upsilon$ might more naturally have been written $\omega \rightarrow
\tau/\nu z \sigma$, $\Upsilon \rightarrow -1/\nu z \sigma$. These 
expressions follow from the $\delta \rightarrow 0$ limit of the
more general scaling hypothesis for the entire distribution close
to the critical point (eq 174). However for this section
we have elected to make
the more restricted hypothesis, eq (53), and to give the exponents
single character names. 

In this section too we neglect injection. At the critical point injection
is a relevant perturbation. The form we derive is therefore a transient
that will break down at sufficient depth. Provided the injection is weak
however that depth could be very great. 

Majumdar and Sire \cite{clement} have analysed the scaling limit of
$\Pi(w,t)$ when injection is present; however it does not appear 
straightforward to take the injection $\rightarrow 0$ limit in
their expression. It would also be desirable for the case of non-zero
injection to have a simple explicit formula for the crossover of
$\Pi(w,t)$ from the transient we derive (eq 53) to the injection
dominated, large depth limit. Presumably this can be accomplished by
extracting the suitable limit of the results of ref \cite{clement},
or by direct calculation, but we do not attempt it here.

\subsection{Disorder Average}

As in section II we assume a uniform load is applied to the top layer
(eq 2). To obtain the distribution $\Pi(w,t)$ following ref \cite{takayasu}
we consider the quantities
\begin{equation}
Z_r (\rho,t) = \langle \exp i \rho \sum_{n=1}^{r} w_n (t) \rangle
\end{equation}
where $r = 1,2,3,\ldots$ By translational invariance 
\begin{eqnarray}
Z_1 (\rho,t) & = & \langle \exp i \rho w_1 (t) \rangle
\nonumber \\
 & = & \sum_{w=0}^{\infty} e^{i \rho w} \Pi (w, t).
\end{eqnarray}
Note that for the critical $q$-model without injection the load on a site
is an integer. Thus $Z_1(\rho,t)$ is the discrete Fourier or $z$-transform
of the distribution $\Pi(w,t)$; $\rho$ is the transform domain variable
conjugate to $w$. $Z_2(\rho,t)$ similarly encodes the joint probability
distribution of load on neighbouring sites and so on. 

For the business at hand the imaginary part of $Z_r(\rho,t)$,
\begin{equation}
{\cal Z}_{r}(\rho,t) = {\rm Im} \hspace{1mm} Z_r(\rho,t),
\end{equation}
is especially valuable. It is evident from eq (55) that
\begin{equation}
{\cal Z}_1 (\rho,t) = \sum_{w=0}^{\infty} \sin (\rho w) \Pi(w,t).
\end{equation}
By using Fourier's identity
\begin{equation}
\frac{2}{\pi} \int_{0}^{\pi} d k \sin k n \sin k m = \delta_{mn}
\end{equation}
and eq (57)
we can extract the distribution $\Pi(w,t)$ from ${\cal Z}_1(\rho,t)$ via
\begin{equation}
\Pi(w,t) = \frac{2}{\pi} \int_{0}^{\pi} d \rho \sin (\rho w) 
{\cal Z}_1 (\rho,t)
\end{equation}
for $w = 1,2,3,\ldots$ We cannot obtain $\Pi(w=0,t)$ in this way from
${\cal Z}_1(\rho,t)$, but we can obtain it from the normalisation of
$\Pi(w,t)$;
\begin{equation}
\Pi(w\rightarrow0,t) = 1 - \sum_{w=1}^{\infty} \Pi(w,t).
\end{equation}

The benefit of considering the quantities $Z_r(\rho,t)$ is that
they obey a simple linear evolution equation. Following ref \cite{takayasu}
write
\begin{eqnarray}
Z_r (\rho,t+1) & = & \langle \exp i \rho \sum_{n=1}^{r} w_n(t+1) \rangle
\nonumber \\
 & = & \langle \exp i \rho \{ w_1 (t) f_1(t) + 
\sum_{n=2}^{r} w_n (t) 
\nonumber \\
& &
+ w_{r+1} (t) [ 1 - f_{r+1}(t) ] \} \rangle.
\end{eqnarray}
To obtain the second line we have used the $q$-model evolution eq (1).
Since the weights in layer $t$ depend only on fractions in the 
preceding layers we can perform the average over $f_1(t)$ and $f_{r+1}(t)$
separately in eq (61):
\begin{eqnarray}
\langle \exp i \rho w_1 (t) f_1 (t) \rangle_{f_1} & = &
\frac{1}{2} \left[ 1 + \exp i \rho w_1 (t) \right];
\nonumber \\
\langle \exp i \rho w_{r+1} (t) [ 1 - f_{r+1} ] \rangle_{f_{r+1}} & = &
\frac{1}{2} \left[ 1 + \exp i \rho w_{r+1} (t) \right].
\end{eqnarray}
Substituting eq (62) in eq (61) we obtain the evolution equation
\begin{equation}
Z_r(\rho,t+1) = \frac{1}{4} Z_{r-1} (\rho,t) + \frac{1}{2} Z_r (\rho,t)
+ \frac{1}{4} Z_{r+1} (\rho,t)
\end{equation}
where we have again made use of horizontal translational invariance.

Note that eq (63) is linear. Hence it is obeyed separately by the
real and imaginary parts of $Z$. ${\cal Z}$ therefore evolves according
to 
\begin{equation}
{\cal Z}_r(\rho,t+1) = 
\frac{1}{4} {\cal Z}_{r-1} (\rho,t) + \frac{1}{2} {\cal Z}_r (\rho,t)
+ \frac{1}{4} {\cal Z}_{r+1} (\rho,t)
\end{equation}
Eq (64) is reminiscent of a tight-binding lattice Schr\"{o}dinger equation
for a free particle on a half-line (since the site index $r \ge 1$).

The main results of this subsection are eqs (57) and (59) that define
the relationship between $\Pi(w,t)$ and ${\cal Z}_1(\rho,t)$ and
eq (64) that controls the evolution of ${\cal Z}_r(\rho,t)$ with depth.

\subsection{Solution and Scaling Limit}

We wish to solve eq (64) subject to the initial condition
\begin{equation}
{\cal Z}_r (\rho, t \rightarrow 0 ) = \sin \rho r.
\end{equation}
This follows from the assumed uniform load applied to the top layer
and eqs (54) and (56). Schematically, eq (64) has the form
\begin{equation}
{\cal Z}_r (\rho,t+1) = \sum_s H_{rs} {\cal Z}_s (\rho,t).
\end{equation}
It is easy to verify that our initial condition is an eigenfunction
of $H$;
\begin{equation}
\sum_s H_{rs} \sin \rho s = \left( \frac{1}{2} + \frac{1}{2} \cos \rho
\right) \sin \rho r.
\end{equation}
Hence eq (64) has the remarkably simple solution
\begin{equation}
{\cal Z}_r (\rho,t) = \left( \frac{1}{2} + \frac{1}{2} \cos \rho
\right)^t \sin \rho r.
\end{equation}
Substituting eq (68) in eq (59) we obtain the desired expression 
for 
\begin{equation}
\Pi (w,t) = \frac{2}{\pi} \int_{0}^{\pi} d \rho \sin ( \rho w )
\left( \frac{1}{2} + \frac{1}{2} \cos \rho \right)^t \sin \rho
\end{equation}
for $w = 1,2,3,\ldots$ 

The integral over $\rho$ can be performed 
exactly by a standard contour integration trick (see footnote 2)
to yield
\begin{eqnarray}
\Pi (w,t) & = & \frac{1}{4^t} \frac{ (2 t)! }{ (t + 1 - w)! (t - 1 + w)! }
\nonumber \\
 & & 
- \frac{1}{4^t} \frac{ (2 t)! }{ (t - 1 - w)! (t + 1 + w)! }
\nonumber \\
& & \hspace{10mm} {\rm for} \hspace{2mm} w = 1,2,\ldots,t-1
\nonumber \\
 & = & \frac{1}{4^t} \frac{ (2 t)! }{ (t + 1 - w)! (t - 1 + w)! }
\hspace{2mm} {\rm for} \hspace{2mm} w = t, t+1
\nonumber \\
 & = & 0 \hspace{5mm} {\rm for} \hspace{2mm} w > t+1.
\end{eqnarray}
We now use eq (60) and (70) to obtain $\Pi(w \rightarrow 0, t)$.
The sum proves tractable and yields
\begin{equation}
\Pi(w \rightarrow 0, t) = 1 - \frac{1}{4^t} \frac{ (2 t + 1)! }{(t+1)! t!}.
\end{equation}
Eq (70) and (71) are the exact expressions for $\Pi(w,t)$ for the critical
$q$-model without injection.

Much more interesting than the exact formula is the scaling limit of
large depth. We now assume $ t \gg 1$ but we will make no assumptions
about the relative size of $w$ and $t$. 
To derive this limit we return to eq (69) and write
\begin{equation}
\left( \frac{1}{2} + \frac{1}{2} \cos \rho \right)^t \approx
e^{- t \rho^2/4}
\end{equation}
justified (inside the integral) for large $t$. Hence we obtain a
Gaussian integral
\begin{eqnarray}
\Pi(w,t) & = & \frac{1}{\pi} \int_{-\pi}^{\pi} d \rho \sin ( \rho w )
\rho e^{- t \rho^2/4}
\nonumber \\
 & = & \frac{4}{\sqrt{\pi}} \frac{w}{t^{3/2}} e^{- w^2/t }.
\end{eqnarray}
Comparing eq (53) and (73) we see that at large depth $\Pi$ has the
anticipated scaling form with exponents
\begin{equation}
\omega = 1, \hspace{3mm} \Upsilon = - \frac{1}{2}
\end{equation}
and scaling function
\begin{equation}
{\cal H}(s) = \frac{4}{\sqrt{\pi}} s e^{-s^2}.
\end{equation}

Eq (73) holds for $w \ge 1$. In the same large depth
limit
\begin{equation}
\Pi (w \rightarrow 0, t) \approx 1 - \frac{2}{\sqrt{\pi}} \frac{1}{\sqrt{t}}.
\end{equation}
The distribution of load thus consists of a spike at zero load followed by 
smooth behaviour for non-zero load given by eq (73). At great depths
it is extremely probable that the load on a given bead is zero; most
of the weight of the distribution is in the spike.

From the distribution of load, eq (73), it is easy to confirm that its
variance (eq 4) grows without bound as the square root of depth, as we had
earlier inferred from the scaling function ${\cal F}$ (cf eq 52).

It is instructive that the exact formula, eqs (70) and (71), is so 
cumbersome; the scaling limit, eqs (53), (74) and (75), emerges only
when we plumb the depths.

\section{Effect of Injection}

In this section we consider the $q$-model in 1+1 dimensions taking
into account injection. We will assume that the weights of the beads
are independent and identically distributed with mean $\langle I 
\rangle$ and variance $ \langle \delta I^2 \rangle$. To probe the
behaviour of the model we will assume that a uniform load is 
applied to the top layer (eq 2). We will study how the mean square
load $ \langle w^2 (t) \rangle $ evolves with depth since the mean 
load has the trivial variation
\begin{equation}
\langle w (t) \rangle = 1 + \langle I \rangle t.
\end{equation}
Near the critical point we expect that the mean square load
should have a scaling form
\begin{equation}
\langle w^2 (t) \rangle = \frac{1}{\delta^{\theta}}
{\cal C} (t \delta^{\varphi}, \langle \delta I^2 \rangle
\delta^{-\mu}, \langle I \rangle \delta^{-\kappa} ).
\end{equation}

We can guess all the exponents and obtain some features of the
scaling function from simple arguments. The load on a particular
bead at depth $t$ is a random linear combination of the weights of
the beads in the layer above plus a term, due to the applied load, 
that does not depend on
the weights, $I_n$. Hence the scaling function has to be of the form
\begin{eqnarray}
\langle w^2 (t) \rangle & = & 
\frac{1}{\delta^{\theta}} {\cal F} (t \delta^{\varphi})
+ \frac{ \langle \delta I^2 \rangle }{ \delta^{\theta + \mu} } 
{\cal M} (t \delta^{\varphi})
+ \frac{ \langle I \rangle }{ \delta^{\theta + \kappa} } 
{\cal K} (t \delta^{\varphi})
\nonumber \\
 & & + \frac{ \langle I \rangle^2 }{ \delta^{\theta + 2 \kappa} }
{\cal L} (t \delta^{\varphi}).
\end{eqnarray}

In the limit of zero injection eq (79) should reduce to our
result in section II. Thus
\begin{equation}
\theta = 1, \hspace{3mm} \varphi = 2
\end{equation}
and ${\cal F}$ has the same form (eq 50) as in section II
justifying the recycling of these particular symbols.

By rewriting the average weight at depth $t$ (eq 77) as 
$ 1 + t \delta^2 \langle I \rangle / \delta^2 $ we conjecture
\begin{equation}
\kappa = 2.
\end{equation}

To obtain $\mu$ we imagine that the system is very close to the
critical point. Then for times that are not too long, effectively,
it will behave as it would right at the critical point. At 
that point the weight of each bead zig-zags down lines that
merge but do not split. If we add the squares of the loads on all
the beads on layer $t$ we will obtain the sum, over all the beads
above layer $t$, of their squared deviation from the average weight
$\langle I \rangle$ plus other terms. Hence 
$ \langle \sum_n w_n (t)^2 \rangle = \langle \delta I^2 \rangle N t +
$ other pieces that do not depend on $ \langle \delta I^2 \rangle $.
Here $N$ is the number of beads in a layer. By translational invariance
we conclude
\begin{equation}
\langle w^2 (t) \rangle \approx \langle \delta I^2 \rangle t +
\hspace{2mm} {\rm others}.
\end{equation}
In eq (82) ``others'' represents contributions to $\langle w^2 (t) \rangle$
that do not depend on $\langle \delta I^2 \rangle$. Comparing eq (82)
and (79) we see that for small values of its argument
\begin{equation}
{\cal M} (u) \approx u
\end{equation}
and the exponent
\begin{equation}
\mu = 1,
\end{equation}
needed to cancel the $\delta$ dependence at small depths

With the exponents in hand we can analyse the behaviour of 
$\langle w^2 (t) \rangle$ at small depths (compared to $1/\delta^2$).
This behaviour would persist out to all depths right at the critical
point. For the term independent of injection we have already obtained
the exact result, eq (52). For the term that depends on $\langle
\delta I^2 \rangle$ we have just worked out the behaviour in this
limit, including the precise coefficient (eq 82). For the term that is
proportional to $\langle I \rangle$ we argue that for small $u$,
${\cal K} (u) \sim u^{3/2}$ to cancel the $\delta$ dependence,
leading to
\begin{equation}
\langle w^2 (t) \rangle \sim \langle I \rangle t^{3/2} + {\rm others}.
\end{equation}
Similarly the contribution of the term that is proportional to 
$\langle I \rangle^2 $ is
\begin{equation}
\langle w^2 (t) \rangle \sim \langle I \rangle^2 t^{5/2} + {\rm others}.
\end{equation}

The last result has a simple interpretation. We have seen in section II
that without injection at the critical point the mean weight at depth
$t$ is 1; the mean square weight $\sim \sqrt{t}$. With injection the
average weight at sufficient depth is $\approx \langle I \rangle t$.
If we assume that uniform injection does not change the distribution,
only its scale, then since the mean is inflated by a factor $\langle
I \rangle t$, the mean square should be inflated by a factor $\langle
I \rangle^2 t^2$, leading to eq (86). The same interpretation can be
used to derive the behaviour of the last term in eq (79)
in the limit $ t \gg 1/\delta^2 $, the opposite of the limit we
have so far considered. In that limit, in the absence of injection, 
the fluctuations saturate at the value $1/\delta$. Hence we expect
this term to behave as
\begin{equation}
\langle w^2 (t) \rangle \sim \frac{1}{\delta} 
\langle I \rangle^2 t^2.
\end{equation}

We can check some of these deductions by making contact with
Majumdar and Sire, who have analysed the entire distribution of
load at the critical point \cite{clement}. Following these authors let us imagine
that the injection term is very small, with the squared mean
$\langle I \rangle^2 $, significantly smaller than the variance
$\langle \delta I^2 \rangle$. According to our analysis ultimately
the fluctuations at the critical point should grow as $t^{5/2}$,
but the depth at which the $t^{5/2}$ term (eq 86) overtakes the
term linear in $t$ (eq 82) could be very great; it diverges as
$1/\langle I \rangle^{4/3}$. Majumdar and Sire arrived at the
same value $4/3$ for this crossover exponent. Moreover, since they
argued that right at the critical point (the only case they considered)
there is only one independent exponent, we have made contact with their
entire analysis as regards exponents. 

In summary we anticipate that near the critical point the mean square
load will follow the scaling form (eq 78). Using simple arguments we
have conjectured values for all the exponents [eqs (80), (81) and (84)]
and guessed some features of the scaling function. As a check we have made
contact with the critical point analysis of Majumdar and Sire and
recovered the known value of the crossover exponent, 4/3 \cite{clement}. 
In the remainder
of this section we will fully confirm the deductions we have made above.
We will obtain an exact formula for the evolution of the mean square
load; the exponents, $\theta, \varphi, \mu $ and $\kappa$; and the scaling
function, ${\cal C}$.

\subsection{Disorder Average and Exact Solution}

As in section II our strategy is to analyse the evolution of the correlation
function, $c_n(t)$; the mean-squared weight $\langle w^2 (t) \rangle = 
c_0 (t) $. The analysis is given in outline since most of the needed
technical elements have already been described in section II. Here we
shall focus on the new complications introduced by consideration of
injection.

Following the method of section IIB we first obtain the evolution equation
for the correlation function, now including injection. Schematically
this equation has the form
\begin{equation}
c_m (t+1) = \sum_n H_{mn} c_n (t) + \xi_m (t).
\end{equation}
$H_{mn}$ is the same matrix as in eq (15). The effect of injection 
appears in the inhomogeneous term $\xi_m$. Explicitly
\begin{equation}
\xi_m = 2 \langle I \rangle + (2 t + 1) \langle I \rangle^2 
+ \langle \delta I^2 \rangle \delta_{m=0}.
\end{equation}

Our strategy to solve eq (88) is to first expand 
$c(t)$ and $\xi(t)$ in terms of the right eigenvectors of $H$:
\begin{equation}
c_m (t) = \sum_{\lambda} a_{\lambda} (t) \phi^{\lambda}_{m};
\hspace{2mm}
\xi_m (t) = \sum_{\lambda} \xi_{\lambda} (t) \phi^{\lambda}_{m}.
\end{equation}
As discussed before, the expansion amplitudes $a_{\lambda}$
and $\xi_{\lambda}$ are calculated by use of the left eigenvectors
\begin{equation}
a_{\lambda}(t) = \sum_m (\psi^{\lambda}_{m})^{*} c_m (t);
\hspace{2mm}
\xi_{\lambda}(t) = \sum_m (\psi^{\lambda}_{m})^{*} \xi_m (t).
\end{equation}
In section IIB we have calculated the amplitudes for $c_m(t \rightarrow 0)$.
We found
$a^{(+)}(k, t \rightarrow 0 ) = a^{(-)}(k, t \rightarrow 0 )
\equiv a (k, t \rightarrow 0)$ with
\begin{equation}
a (k, t \rightarrow 0) = 2 \pi {\cal A} (k)^{*} \delta (k)
+ [ 1 - {\cal A}(k)^{*} ] i \cot \frac{k}{2}.
\end{equation}
Here ${\cal A}(k)$ is given by eq (35). Similarly
$\xi^{(+)}(k, t \rightarrow 0 ) = \xi^{(-)}(k, t \rightarrow 0 )
\equiv \xi (k, t \rightarrow 0)$ with
\begin{eqnarray}
\xi (k,t) & = & \langle \delta I^2 \rangle {\cal A}(k)^{*}
+ \{ 2 \langle I \rangle + (2 t + 1) \langle I \rangle^2 \}
\nonumber \\
& &
\{ 2 \pi {\cal A}(k)^{*} \delta (k) + 
[1 - {\cal A}(k)^{*} ] i \cot \frac{k}{2} \}.
\end{eqnarray}

Substituting the expansions eq (91) into the evolution eq (88) shows that the
dynamics of the amplitudes for different right eigenvectors is decoupled
and is given by
\begin{equation}
a_{\lambda} (t + 1) = \lambda a_{\lambda} (t) + \xi_{\lambda} (t).
\end{equation}
To solve this dynamics we introduce the $z$-transforms
\begin{eqnarray}
a_{\lambda} (z) & = & \sum_{t=0}^{\infty} a_{\lambda} (t) z^t,
\nonumber \\
\xi_{\lambda} (z) & = & \sum_{t=0}^{\infty} \xi_{\lambda} (t) z^t,
\end{eqnarray}
to obtain
\begin{equation}
a_{\lambda} (z) = \frac{ a_{\lambda} (t \rightarrow 0) }{1 - z \lambda}
+ \frac{z \xi_{\lambda} (z) }{1 - z \lambda}.
\end{equation}

Combining eq (90) and (96) we conclude
\begin{equation}
c_m (z) = \sum_{\lambda} \left[ \frac{ a_{\lambda} (t \rightarrow 0) }{1 -
z \lambda} + \frac{ z \xi_{\lambda} (z) }{ 1 - z \lambda }
\right] \phi^{\lambda}_{m}.
\end{equation}
More explicitly
\begin{eqnarray}
c_{0} (z) & = & 2 \int_{0}^{\pi} d k 
\frac{ a(k, t \rightarrow 0) A(k) }{1 - z \lambda(k)}
\nonumber \\
 & & + 2 z \int_{0}^{\pi} d k
\frac{ \xi(k, z) A(k) }{1 - z \lambda(k)}.
\end{eqnarray}
Here $c_0 (z)$ is the $z$-transform of $c_0(t)$; $A(k)$ is given by
eq (30); $\lambda(k)$, by eq (29); and $a(k, t \rightarrow 0)$, by eq (92).
$\xi(k,z)$ is to be obtained by $z$-transforming eq (93). 

Now all the pieces have been assembled. It remains to perform the
$k$ integral and invert the $z$-transform. The $k$-integrals may be
performed by the standard contour integration method mentioned in 
footnote 2. The $z$-transforms can all be inverted as illustrated in
Appendix B. 

After much calculation we find
\begin{equation}
\langle w^2 (t) \rangle = \overline{F} (t, \epsilon)
+ \langle \delta I^2 \rangle M (t, \epsilon)
+ \langle I \rangle K (t, \epsilon)
+ \langle I \rangle^2 L (t, \epsilon)
\end{equation}
with
\begin{eqnarray}
\overline{F} (t, \epsilon) & = & \frac{\epsilon}{1 - \epsilon}
- \frac{1}{\pi \epsilon} \Gamma F_1;
\nonumber \\
M (t, \epsilon) & = & - \frac{\epsilon}{ (1 - \epsilon)^2 }
+ \frac{2}{\pi} \frac{(1 - \epsilon)}{\epsilon^2} \Gamma [ t F_1 + F_2 ];
\nonumber \\
K (t, \epsilon) & = & \frac{2}{1 - \epsilon} t +
\frac{2 \epsilon^2}{ (1 - \epsilon)^3 } 
- \frac{4}{\pi \epsilon} \Gamma [ t F_1 + F_2 ];
\nonumber \\
L (t, \epsilon) & = & \frac{\epsilon^4 + 2 \epsilon^3 - \epsilon^2}{ (1 - 
\epsilon)^5 }
+ \frac{ 2 \epsilon^2 }{ (1 - \epsilon)^3 } t
\nonumber \\
 & & + \frac{2}{3 \pi \epsilon} \Gamma [ (4 t^2 - t) F_1
 + (8t - 5) F_2 + 8 F_3 ].
\end{eqnarray}
We have put an overline on $\overline{F} (t,\epsilon)$ to avoid confusion
with a hypergeometric function. For brevity we have written
\begin{eqnarray}
\Gamma & = & \frac{ \Gamma ( 1/2 ) \Gamma( t + 3/2 ) }{ \Gamma(t + 2) };
\nonumber \\
F_n & = & 
F \left( n, t + \frac{3}{2}, t + 2; \frac{ 2 \epsilon - 1 }{\epsilon^2} \right)
\end{eqnarray}
in eq (100).

Eq (100) is the final result of this subsection. It gives the evolution
of load fluctuations for the $q$-model with injection in 1+1 dimensions.
It holds for any distribution of fractions $P(f)$ and at any depth.

\subsection{Scaling Limit}

More interesting than the exact results is the scaling behaviour
that emerges for $t \gg 1$ and $\delta = 1 - \epsilon \rightarrow
0$. To derive this behaviour it is useful to express the hypergeometric
functions in eq (100) via the integral representation, eq (45). The asymptotic
behaviour of $\Gamma F_1$ has been analysed in section IIC [cf. eqs (47)
and (48)]. The corresponding analysis of $\Gamma F_2$ and $\Gamma F_3$
is very similar and finally leads to
\begin{eqnarray}
\overline{F} (t, \epsilon) & \rightarrow &
\frac{1}{\delta}
\left\{ 1 - \frac{2}{\pi} \Phi_{1} (u) \right\};
\nonumber \\
M(t, \epsilon) & \rightarrow & 
\frac{1}{\delta^2} \left\{ -1 + \frac{4}{\pi} u \Phi_1(u) + 
\frac{4}{\pi} \Phi_2 (u) 
\right\};
\nonumber \\
K(t, \epsilon) & \rightarrow & 
\frac{1}{\delta^3} 
\left\{2 u + 2 - \frac{8 u}{\pi} \Phi_1(u) - \frac{8}{\pi} \Phi_2(u)
\right\};
\nonumber \\
L(t, \epsilon) & \rightarrow & 
\frac{1}{\delta^5}
\{ 2 + 2 u + u^2
\nonumber \\
 & & - \frac{16}{3 \pi} \left[ u^2 \Phi_1(u) + 2 u \Phi_2 (u) + 
 2 \Phi_3 (u) \right] \}.
\end{eqnarray}
Here $u = t \delta^2$. For brevity we have written
\begin{equation}
\Phi_n (u) = \int_{0}^{\infty} d s \frac{ e^{-u s^2} }{ (1 + s^2)^n }.
\end{equation}
Comparing eq (79) to (102) we conclude that the exponents are 
$\theta = 1, \varphi = 2, \kappa = 1$ and $\mu = 1$ as conjectured.
It is also straightforward to extract the scaling functions
${\cal F}(u), {\cal M}(u), {\cal K}(u)$ and ${\cal L}(u)$ 
from eq (102). The scaling functions are plotted in figs 4, 5, 6
and 7 respectively.

The asymptotics of the integrals $\Phi_n (u)$ are analysed in
Appendix C. Using those results we conclude that for small $u$
\begin{eqnarray}
{\cal F}(u) & \approx & \frac{2}{\sqrt{\pi}} u^{1/2};
\nonumber \\
{\cal M}(u) & \approx & u;
\nonumber \\
{\cal K}(u) & \approx & \frac{8}{3 \sqrt{\pi} } u^{3/2};
\nonumber \\
{\cal L}(u) & \approx & \frac{16}{15 \sqrt{\pi} } u^{5/2}.
\end{eqnarray}
For large $u$
\begin{eqnarray}
{\cal F}(u) & \approx & 1;
\nonumber \\
{\cal M}(u) & \approx & \frac{2}{\sqrt{\pi}} \sqrt{u};
\nonumber \\
{\cal K}(u) & \approx & 2 u;
\nonumber \\
{\cal L}(u) & \approx & u^2.
\end{eqnarray}

Substituting the small $u$ asymptotics in eq (79) we obtain
the behaviour for depths small compared to $1/\delta^2$.
\begin{equation}
\langle w^2 (t) \rangle \approx \frac{2}{\sqrt{\pi}} t^{1/2}
+ \langle \delta I^2 \rangle t 
+ \langle I \rangle \frac{8}{3 \sqrt{\pi}} t^{3/2}
+ \langle I \rangle^2 \frac{16}{15 \sqrt{\pi}} t^{5/2}.
\end{equation}
This behaviour would persist for all depths right at the critical
point. Note that eq (106) agrees with the forms conjectured in
eq (82), (85) and (86) (including the numerical coefficient in the first
case). It is hardly necessary to add that eq (106) is consistent
with the critical point analysis of Majumdar and Sire since it leads,
by the arguments given earlier in this section, to their crossover
exponent 4/3 \cite{clement}. 

The large $u$ asymptotics give the behaviour at depths large 
compared to $1/\delta^2$. We find
\begin{equation}
\langle w^2 (t) \rangle = 
\frac{1}{\delta} 
+ \langle \delta I^2 \rangle \frac{1}{\delta} \frac{2}{\sqrt{\pi}} t^{1/2}
+ \langle I \rangle \frac{1}{\delta} 2 t
+ \langle I \rangle^2 \frac{1}{\delta} t^2.
\end{equation}
The term proportional to $\langle I \rangle^2$ has the form 
anticipated in eq (87); at the greatest depths this term is dominant.

In summary we have shown that the singular distribution is an isolated
critical point. Near the critical point the fluctuations in load
have the scaling form eq (79). We have derived this scaling form
and all the exponents. The results are in agreement with expectations
based on simpler (non-rigorous) arguments.

\section{Higher Dimensions}

We now turn to the $q$-model in $D+1$ dimensions. The quantum
Hall multilayer and river networks are both 1+1 dimensional
systems; bead-packs however are described by the 2+1 dimensional
$q$-model. The behaviour of the model as a function of $D$ is of
intrinsic interest moreover. We will find that right at the critical
point the growth exponents vary smoothly with dimension for $D <2$.
Above $D=2$ they become fixed, revealing $D=2$ as the upper critical
dimension for the critical case. Off the critical point we expect the 
fluctuations to grow according to a scaling function ${\cal F}(x)$ (eq 7).
We will study how the function and exponents vary with dimensionality
below $D=2$. For simplicity in this section
we neglect injection.

\subsection{Model and Disorder Average}

First we must generalise the description of the $q$-model, so far
confined to 1+1 dimensions. The case of 2+1 dimensions is easy to
visualise. Fig 8 illustrates a square lattice composed of two interpenetrating
square sublattices. The co-ordinates of sites $\vec{n} = (n_1, n_2)$ are both
even for the black sites; both odd for the grey. The displacements from a site
on either sublattice to its four nearest neighbours on the {\em other} 
sublattice are $ (\pm 1, \pm 1)$. We will denote these displacements
$\vec{u}$. In the $q$-model planes of such square 
lattices are stacked vertically.
The beads alternately occupy only even or odd sublattices. Denoting the 
depth of a layer $t$, for $t$ even only the even sublattice is occupied;
for $t$ odd, only the odd sublattice. Viewed in three dimensions the 
beads occupy a body-centered cubic stucture. In the same sense, Fig 1
can be viewed as a body-centered square structure. 

Now consider a $D$ dimensional simple cubic lattice. The co-ordinates
of a site are specified by $\vec{n} = (n_1, n_2, \ldots, n_D)$ 
where $n_i$ are integers. For the even sublattice the $n_i$ are even;
for the odd sublattice, they are odd. Each site has $2^D$ nearest
neighbours on the other sublattice. We denote the displacements
$(\pm 1, \pm 1, \ldots, \pm)$ to these neighbours $\vec{u}$. The $D+1$
dimensional $q$-model consists of $D$ dimensional cubic lattices stacked
in the ``vertical'' $t$ direction. In alternate $t$ slices only the even
or odd sublattices are occupied by beads. 

It is assumed that a random fraction of the load on each bead is
supported by its $2^D$ neighbours in the layer below. The fractions
must sum to one;
\begin{equation}
\sum_{\vec{u}} f_{\vec{u}} = 1.
\end{equation}
Here $f_{\vec{u}}$ is the fraction of load transmitted by the bead
to the neighbour separated by a horizontal displacement of $\vec{u}$.
Hence the dynamics of the $q$-model is governed by 
\begin{equation}
w (\vec{n}, t+1) = \sum_{\vec{u}} f_{\vec{u}} ( \vec{n} - \vec{u}, t )
w (\vec{n} - \vec{u}, t).
\end{equation}
Eq (109) is the $D+1$ dimensional generalisation of eq (1).

The fractions for a particular bead are assumed to be drawn from
a distribution that is symmetric with respect to direction
and respects the constraint eq (108).
It follows
\begin{equation}
\langle f_{\vec{u}} \rangle = \frac{1}{2^D}.
\end{equation}
We write 
\begin{equation}
\langle f_{\vec{u}}^2 \rangle = \frac{1}{2^{2D}} + \frac{\epsilon}{2^{2D}}
\end{equation}
where $\epsilon$ is a parameter that characterises the distribution of
fractions. From the sum constraint eq (108) it follows
\begin{equation}
\langle f_{\vec{u}_1} f_{\vec{u}_2} \rangle = \frac{1}{2^{2D}}
- \frac{\epsilon}{(2^D - 1)} \frac{1}{2^{2D}}.
\end{equation}
for $\vec{u}_1 \neq \vec{u}_2$.
The fractions are assumed to be independently and identically distributed
for different beads. 

For the singular distribution all the fractions are zero except one.
The probability for each fraction to be one is $1/2^D$. It is easy
to calculate $\epsilon = 2^D - 1$ for the singular distribution using
eq (111) and to verify eq (110) and (112) are satisfied. 
Since $\epsilon = 2^D - 1$ for the singular distribution we shall use
$\delta$, defined by
\begin{equation}
\delta = 1 - \frac{\epsilon}{(2^D - 1)},
\end{equation}
as our measure of the distance of a distribution from the critical point.

As before it is useful to consider the correlation function
\begin{equation}
c(\vec{m},t) = \sum_{\vec{n}} \langle w (\vec{n},t) w (\vec{n} + \vec{m}, t)
\rangle.
\end{equation}
Note that $\vec{m}$ is a $D$ dimensional vector with even integer entries
for both $t$ even and $t$ odd. The correlation function therefore lives on
a simple cubic lattice in $D$ dimensions. By rescaling, as in section IIA,
we reduce the lattice constant of this lattice to one so that the components
of the vector $\vec{m}$ are now integers. The variance in load is related
to the on-site correlation by
\begin{equation}
\langle \delta w^2 (t) \rangle = c (\vec{m} \rightarrow 0, t) - 1.
\end{equation}

Following the discussion of section IIA and using eqs (109), (110),
(111) and (112) it is easy to show that the correlation function 
evolves with depth according to
\begin{eqnarray}
c (\vec{m}, t+1) & = & \frac{1}{2^D} c(\vec{m},t) + \frac{1}{2^{D+1}}
\sum_{\vec{b} = {\rm nn}} c( \vec{m} + \vec{b}, t)
\nonumber \\
 & & + \frac{1}{2^{D+2}} \sum_{\vec{b} = {\rm nnn}} c( \vec{m} + \vec{b}, t)
+ \ldots 
\nonumber \\
& &
+ \frac{1}{2^{2D}} \sum_{\vec{b} = {\rm n \ldots n}} 
c( \vec{m} + \vec{b}, t) + \frac{\epsilon}{2^D} c (\vec{m},t) 
\delta_{\vec{m}=0}
\nonumber \\
& &
- \frac{\epsilon}{(2^D - 1)} \frac{1}{2^{D+1}}
\sum_{\vec{b} = {\rm nn}} c( \vec{m} + \vec{b}, t) 
\delta_{\vec{m} + \vec{b} = 0}
\nonumber \\
& &
- \frac{\epsilon}{(2^D - 1)} \frac{1}{2^{D+2}}
\sum_{\vec{b} = {\rm nnn}} c( \vec{m} + \vec{b}, t) 
\delta_{\vec{m} + \vec{b} = 0}
\nonumber \\
& & 
\ldots - \frac{\epsilon}{(2^D - 1)} \frac{1}{2^{2D}}
\sum_{\vec{b} = {\rm n \ldots n}} c( \vec{m} + \vec{b}, t)
\delta_{\vec{m} + \vec{b} = 0}
\end{eqnarray}
While reading eq (116) it is useful to recall that the correlation 
function lives on a $D$-dimensional cubic lattice. For $D=2$ each site
has four nearest neighbours and four next-nearest neighbours. For general
$D$, each site has $2D$ nearest neighbours; $ 2^2 C(D,2) $ next nearest
neighbours; $2^3 C(D,3) $ third nearest neighbours; and $2^D C(D,D) $
$D^{\rm th}$ nearest neighbours. In eq (116) $\vec{b}$ denotes the displacement
from a site to any of these neighbours; nn denotes nearest neighbour; nnn,
next nearest; and so forth.

In the next subsection we will solve eq (116) for $c(\vec{m} \rightarrow 0,
t)$ subject to the initial condition that a uniform load has been
applied to the top layer. Thus $ c( \vec{m}, t \rightarrow 0) = 1$
for all $\vec{m}$. 

\subsection{Solution}

It is easy to verify that 
\begin{eqnarray}
c(\vec{m},t \rightarrow \infty) & = & \frac{1}{\delta} \hspace{2mm}
{\rm for} \hspace{2mm} \vec{m} = 0,
\nonumber \\
& & = 1 \hspace{2mm} {\rm otherwise}, 
\end{eqnarray}
is a steady state solution to eq (116). Eq (117) shows that the 
variance $\langle \delta w^2 \rangle$ saturates at sufficient depth
in all dimensions for all distributions except the singular.

We now calculate the evolution of the variance with depth using a method
different from that of section II \cite{yikuo}. 
First we $z$-transform the (discrete)
$t$ dependence of the correlation function,
\begin{equation}
c (\vec{m},z) = \sum_{t=0}^{\infty} c( \vec{m}, t) z^t,
\end{equation}
and Fourier transform the space dependence,
\begin{equation}
c (\vec{p}, z) = \sum_{m} e^{- i \vec{p}.\vec{m} }c( \vec{m}, z).
\end{equation}
The use of the same symbol for the correlation and its transforms,
although customary, is potentially confusing. For example, $c(\vec{p},t
\rightarrow 0) $ denotes the Fourier transform of $c(\vec{m},t)$ at $t=0$;
no $z$-transform is implied. 

Performing both transforms on eq (116) we obtain
\begin{eqnarray}
c (\vec{p},z) & = & c (\vec{p}, t \rightarrow 0) + z c(\vec{p},z) S(\vec{p})
\nonumber \\
& & + \frac{\epsilon}{2^D - 1} z c(\vec{m} \rightarrow 0, z).
\end{eqnarray}
Here
\begin{eqnarray}
S(\vec{p}) & = & \frac{1}{2^D} \{ 1 + \frac{1}{2} \sum_{b = {\rm nn}}
e^{i \vec{p}.\vec{b}} + \frac{1}{2^2} \sum_{b = {\rm nnn}}
e^{i \vec{p}.\vec{b}} 
\nonumber \\
& & + \ldots + \frac{1}{2^D} \sum_{b = {\rm n \ldots n}}
e^{i \vec{p}.\vec{b}} \}
\nonumber \\
& = & \frac{(1 + \cos p_1)}{2} \frac{(1 + \cos p_2)}{2} \ldots
\frac{(1 + \cos p_D)}{2}
\end{eqnarray}
is a ``structure factor'' for the cubic lattice. It will also
prove convenient to define
\begin{equation}
G(\vec{p},z) = \frac{1}{1 - z S(\vec{p})}.
\end{equation}
Both $S(\vec{p})$ and $G(\vec{p},z)$ have helpful physical interpretations
that we shall make use of below. 
For the moment we rearrange eq (120) to obtain 
\begin{eqnarray}
c(\vec{p},z) & = & c( \vec{p}, t \rightarrow 0 ) G(\vec{p},z) 
\nonumber \\
& &
+ (1 - \delta) z c(\vec{m} \rightarrow 0, z) [ 1 - S(\vec{p}) ]
G(\vec{p},z).
\end{eqnarray}
By inverting the Fourier transform we can turn eq (123) into an
expression for $ c(\vec{m} \rightarrow 0, z)$. After further 
re-arrangement
\begin{eqnarray}
c(\vec{m} \rightarrow 0, z) & = & \frac{ \int \frac{d \vec{p}}{ (2 \pi)^D }
c (\vec{p}, t \rightarrow 0) G(\vec{p}, z) }{ 1 - (1 - \delta) z
\int \frac{d \vec{p}}{ (2 \pi)^D } [ 1 - S(\vec{p}) ] G( \vec{p}, z ) }.
\end{eqnarray}
Eq (124) is a general expression for $c(\vec{m} \rightarrow 0, z)$ for
an arbitrary initial condition. For uniform loading of the top layer
\begin{equation}
c (\vec{p}, t \rightarrow 0) = (2 \pi)^D \delta (\vec{p}).
\end{equation}
It follows from eq (121) and (122) that 
$G(\vec{p} \rightarrow 0, z) = 1/(1 - z)$; hence eq (124) simplifies to
\begin{eqnarray}
& & c ( \vec{m} \rightarrow 0, z )  =  (1 - z)^{-1} 
\nonumber \\
& & \times
\left\{ 1 - (1 - \delta) z \int \frac{d \vec{p}}{(2 \pi)^D}
[ 1 - S(\vec{p}) ] G(\vec{p}, z) \right\}^{-1}. 
\end{eqnarray}
Eq (126), together with the definitions of the structure factor
(eq 121) and $G(\vec{p},z)$ (eq 122), constitutes an exact
formal evaluation of the variance with depth. To obtain 
$\langle \delta w^2 (t) \rangle$ explicitly it only remains to
peform the integral over $\vec{p}$ and to invert the $z$-transform.
We return to this task in the next subsection. We conclude this
subsection with a useful interpretation of $S(\vec{p})$ and
$G(\vec{p},z)$.

Eq (116) with $\epsilon \rightarrow 0$ resembles the Schr\"{o}dinger
equation for a particle on a $D$-dimensional cubic lattice with hopping
to the nearest neighbours, the next nearest neighbours, and so on to
the $D^{\rm th}$ nearest neighbours. It is not difficult to see that the
eigenstates of this Schr\"{o}dinger equation are plane waves. $S(\vec{p})$
is the dispersion relation, the eigenvalue at wave vector $\vec{p}$. 
From eq (121) we see that the energy level spectrum is a continuous
band between zero and one. 

The momentum space Green's function for this tight-binding lattice 
would normally be written
\begin{equation}
{\cal G} (\vec{p},E) = \frac{1}{E - S(\vec{p})}.
\end{equation}
Comparing eq (127) to 
eq (122) we see that $G(\vec{p},z)$ is essentially the Green's function
with $E \rightarrow 1/z$.
It is familiar from quantum mechanics that the real space Green's
function at the origin,
\begin{equation}
{\cal G} ( \vec{m} \rightarrow 0, E ) =
\int \frac{d \vec{p}}{(2 \pi)^D} \frac{1}{E - S(\vec{p})},
\end{equation}
regarded as a function of (complex) $E$, has a branch cut
running from $E = 0$ to $E = 1$, the interval that 
supports the eigenvalue band.  
It is not difficult to use the familiar
arguments to conclude that, regarded as a function of complex $z$,
$c( \vec{m} \rightarrow 0, z)$ has a branch cut along the line $z=1$
to $\infty$ (onto which the segment [0,1] maps under the transformation
$E \rightarrow 1/z$). The analytic properties of $c(\vec{m}
\rightarrow 0,z)$ will prove useful in the next sub-section.

\subsection{Scaling Limit}

In this subsection we study the evolution of the variance in the large depth
scaling limit. 
Thus $t \gg 1$ and $\delta$ is zero or very close to it throughout.

An advantage of studying the large depth limit is that we do not have
to calculate $c(\vec{m} \rightarrow 0, z)$ exactly; it is only necessary
to calculate the leading behaviour as $z \rightarrow 1$.
One way to understand this is to consider the critical case 
$\delta = 0$. In this case we expect that at great depth
\begin{equation}
c (\vec{m} \rightarrow 0, t) \sim t^{x}.
\end{equation}
It is easy to show that for $f(t) = t^{x}$, the $z$-transform is
$\Gamma( x + 1 )/(1 - z)^{x + 1}$ 
plus less singular terms. Thus for a function that
behaves as $t^{x}$ for large $t$ also the $z$-transform is
\begin{equation}
t^{x} \leftrightarrow \frac{\Gamma( x + 1)}{(1 - z)^{x+1}} +
\hspace{2mm} {\rm less} \hspace{2mm} {\rm singular}.
\end{equation}
If we know the leading singularity of $c(\vec{m} \rightarrow 0, z)$
as $z \rightarrow 1$ we can use eq (130) to read off the large depth
behaviour.

Another way to see that we only need the behaviour of $c(\vec{m} \rightarrow
0, z)$ as $z \rightarrow 1$ is to consider inverting the $z$-transform
by the contour integral method of Appendix B. This is accomplished by
folding the contour over the branch point of $c(\vec{m} \rightarrow 0, z)$
at $z=1$ and integrating along the cut. In that integral $c(\vec{m} 
\rightarrow 0, z)$ is weighted by a factor that decays extremely
rapidly away from $z=1$ at large depths.

Our goal therefore is to analyse the $z \rightarrow 1$ behaviour of 
\begin{equation}
G(z) = \int \frac{d \vec{p}}{ (2 \pi)^D } \frac{1}{1 - z S(\vec{p})}
\end{equation}
since by a straightforward re-arrangement the integral in eq (126)
simplifies to
\begin{equation}
\int \frac{d \vec{p}}{ (2 \pi)^D } [1 - S(\vec{p})] G(\vec{p},z)
= \left( 1 - \frac{1}{z} \right) G(z) + \frac{1}{z}.
\end{equation}
Insight into the behaviour of $G(z)$ can be gained by expanding
$S(\vec{p})$ around $\vec{p}=0$ to obtain
\begin{equation}
G(z) \approx \int \frac{d \vec{p}}{ (2 \pi)^D } 
\frac{1}{(1 - z) + \vec{p}^2 / 4}.
\end{equation}
If we set $z=1$ in eq (133) the integrand diverges as $\vec{p} \rightarrow
0$ for $D \leq 2$; it is regular in more than two dimensions. Thus in more
than two dimensions $G(z)$ has a branch point at $z=1$ but there is no
actual divergence. In two dimensions or less there is an actual divergence.

The leading behaviour of $G(z)$ above two dimensions is thus simply obtained
by setting $z=1$ in eq (131):
\begin{equation}
G(z) \approx G(1) \hspace{2mm} {\rm for} \hspace{2mm} D > 2.
\end{equation}
In two dimensions we can obtain the singularity by recognising 
$G(z)$ to be a Jacobi elliptic integral. Square lattice Green's 
functions are known to be related to Jacobi's elliptic functions;
but since our lattice features next-nearest neighbour hopping,
in addition to the customary nearest neighbour hopping, we outline
the analysis in Appendix D. The result is that for $z \rightarrow 1$
\begin{equation}
G(z) \approx - \frac{1}{\pi} \ln (1 - z) \hspace{2mm}
{\rm for} \hspace{2mm} D=2.
\end{equation}
For $D < 2$ we obtain the singular behaviour of $G(z)$ in
Appendix D. The result is
\begin{equation}
G(z) = \frac{ \Gamma(1 - D/2) }{\sqrt{\pi}^D} (1 - z)^{D/2 - 1}
\hspace{2mm} {\rm for} \hspace{2mm} D < 2.
\end{equation}
An important feature revealed by this calculation is that the singular
behaviour of $G(z)$ is controlled by the long wavelength behaviour
of $G(p,z)$ for all $D < 2$; it breaks down as $D \rightarrow 2$. 
Although it is instructive to do the calculation for continuous
$D$ to examine the $D \rightarrow 2$ limit, the only case that
is physically relevant is of course the integer dimension $D=1$.

Equipped with the leading behaviour of $G(z)$ in all dimensions
we now obtain the long time behaviour of $\langle \delta w^2 (t)
\rangle $. At the critical point we set $\delta = 0$ and substitute
eqs (132), (134), (135) and (136) in eq (126). Except in two dimensions
the $z$-transforms may be inverted by inspection of eq (130). For
two dimensions we must resort to the method of Appendix B and finally
obtain
\begin{eqnarray}
\langle \delta w^2 (t) \rangle & = & \pi^{D/2 -1 }
\sin \left( \frac{\pi D}{2} \right) t^{D/2} 
\hspace{2mm} {\rm for} \hspace{2mm} D < 2
\nonumber \\
& = & \frac{t}{\ln t} \hspace{2mm} {\rm for} \hspace{2mm} D = 2
\nonumber \\
& = & \frac{1}{G(1)} t \hspace{2mm} {\rm for} \hspace{2mm} D > 2.
\end{eqnarray}
As indicated by the simple steady state solution, at the critical
point the fluctuations
grow without bound as a power of $t$ for all dimensions. The exponent
becomes independent of $D$ for $D > 2$ revealing $D=2$ as the upper
critical dimension.

By substituting eq (132) and (136) in eq (126) we can also obtain
the behaviour of $\langle \delta w^2 (t) \rangle$ away from the
critical point for less than two dimensions. Inverting the $z$-transform
by the method of Appendix B we find
\begin{equation}
\langle \delta w^2 (t) \rangle = \frac{1}{\delta^{\theta}}{\cal F}
(t \delta^{\varphi}).
\end{equation}
Here the exponents
\begin{equation}
\theta = 1, \hspace{2mm} \varphi = \frac{2}{D}
\end{equation}
and the scaling function 
\begin{equation}
{\cal F} (u) = \frac{1}{D} - \frac{2}{\pi D} 
\int_{0}^{\infty} d s \frac{1}{1+ s^2} \exp - \frac{u s^{2/D}}{q_D^{2/D}}
\end{equation}
with $q_D = \Gamma(1 - D/2) \sin ( \pi D/2) /\sqrt{\pi}^D$ a dimension
dependent constant. Again, only the result for $D=1$ is
physically meaningful; in this case eq (140) coincides with the
result of section II. 

In summary the main results of this section are that for all distributions,
except the singular, at sufficient depth the load fluctuations saturate
and (in agreement with experiment) there are no horizontal correlations
in load (eq 117). The saturation value of the load variance diverges
as the critical point is approached. At the critical point the load
fluctuations grow without bound as a power of depth (eq 137). Below
two dimensions this exponent depends on dimensionality; above two
dimensions it is constant, revealing $D=2$ as the critical dimension.
At the critical dimension the growth of fluctuations is tempered by
a logarithmic factor as might be expected at a critical dimension. 
We have also evaluated the scaling function that describes the growth
and saturation of load fluctuations near the critical point for
$D < 2$.

\section{Quantum Hall Multilayer}

\subsection{Models}

In this section we turn to the chiral wave models that are 
believed to adequately describe the surface electronic states
of a quantum Hall multilayer. We begin by examining the circumstances
under which the quantum network model of Saul, Kardar and Read 
\cite{saul}
discussed in section I is equivalent to a $q$-model. 

Following Saul, Kardar and Read, the first step is to identify
pairs of links (joined by vertical
grey bars in Fig 9) as ``beads''. The ``load'' on a bead is the total 
probability that the electron is on either of its two constituent
links. Load propagates from left to right now rather
than top to bottom as it did in our earlier depictions of the $q$-model.
For this reason we will label the vertical co-ordinate $n$ and the
horizontal co-ordinate $t$ here (see fig 3). 

To analyse how load propagates consider an elementary vertex of
the Saul, Kardar and Read model shown in Fig 9. The wave function
amplitudes are related via
\begin{equation}
\left( \begin{array}{c}
\phi_2 \\
\phi_3 
\end{array}
\right)
= S 
\left( \begin{array}{c}
\psi_1 \\
\psi_2
\end{array}
\right);
\end{equation}
here $S$ is a random $2 \times 2$ su(2) rotation matrix.
Saul, Kardar and Read assumed the $S$-matrices were drawn
from the invariant distribution for the su(2) group \cite{hamermesh}. The loads
on beads A, B and C are respectively $|\psi_1|^2 + |\psi_2|^2$,
$|\phi_1|^2 + |\phi_2|^2$ and $|\phi_3|^2 + |\phi_4|^2$ . By unitarity
$|\psi_1|^2 + |\psi_2|^2 = |\phi_3|^2 + |\phi_4|^2$. Thus bead A
sends a fraction $f$ of its load to neighbour B and the remainder
$1-f$ to neighbour C. 

A key feature of the Saul, Kardar and Read model is that the distribution
of the fractions, $P(f)$ is independent of the input amplitudes $\psi_1$ 
and $\psi_2$. This follows from the assumed group invariant distribution
for the $S$-matrices. It is this feature that allows the Saul, Kardar
and Read model to be mapped onto a $q$-model. 

To derive the distribution of the fractions recall that an su(2) matrix
may be parametrized $S = x_0 + i \vec{x}. \vec{\sigma}$ with
$(x_0, \vec{x})$ real and subject to $x_0^2 + \vec{x}^2 = 1$. If
we take $\psi_1 = 1, \psi_2 = 0$ then $f = x_0^2 + x_1^2$. From
the invariant distribution for su(2) matrices,
\begin{equation}
P(x_0, \vec{x}) = \frac{1}{\pi} \delta ( x_0^2 + \vec{x}^2 - 1),
\end{equation}
it is not difficult to show that the fraction $f$ follows the uniform
distribution, $P(f) = 1$ for $0 < f < 1$. 

Now suppose the wave function is known through the vertical slice
$t=0$. We could propagate the wave function $t$ slices to the right
using the quantum Saul, Kardar and Read model. Alternatively we could
calculate the load in the initial layer and propagate it to the right
using the $q$-model with uniform distribution. Either way the load we obtain
in layer $t$ would be the same statistically. This is the sense in which
the Saul, Kardar and Read model is equivalent to the $q$-model.

Note that the $q$-model does not keep track of phase information. The
mapping is useful only under circumstances that the phase information 
is unimportant. Below we will discuss some problems of wave packet
dynamics for which the mapping is useful. The mapping can also be used
to study vertical transport in the quantum Hall multilayer in the limit
of large circumference but we do not discuss that application here.

An obvious circumstance when the phase information is important
and the mapping cannot be used is if periodic boundary conditions
are imposed in the horizontal $t$-direction, as would be appropriate
for a multilayer in the fully phase-coherent, mesoscopic regime.
Phase information is needed to match the wavefunction after it is
propagated around the circumference. We will develop this point in a
more technical way in subsection VI C. 

Another case in which a quantum network model will map onto a $q$-model
is if the wavefunctions and $S$-matrices are chosen to be real and the 
$S$-matrices are further assumed to be distributed over the subgroup of
rotations about the $y$-axis with appropriate invariant measure. The fraction
distribution $P(f) = (1/\pi) f^{-1/2} (1 - f)^{-1/2}$ for the $q-$model
that results. For most distributions of the $S$-matrix however it is not
possible to obtain even the limited mapping between the quantum network
model and the classical $q$-model obtainable in this and in
the Saul, Kardar and Read case.

Finally we present a convenient continuum model of the multilayer surface
governed by the Schr\"{o}dinger equation
\begin{equation}
- i \frac{\partial}{\partial t} \psi_n (t) = 
m_n (t) \psi_{n+1} (t) + m_{n-1}^{*} (t) \psi_{n-1} (t).
\end{equation}
Since the equation is first order in $t$, given the wavefunction at a
fixed $t$ slice we can use it to propagate the wavefunction to the
right, just as in the discrete network model. In the transverse direction
the model is discrete and second -order. Disorder is incorporated
by taking the hopping elements $m_n(t)$ to be random. For a
discussion of the relationship between onsite and hopping disorder
see ref \cite{heisenberg,zirnbauer}. Evidently this model cannot be
reduced to a classical $q$-model.

\subsection{Wave-packet dynamics}

In this section we briefly discuss wave packet dynamics
for the models of the previous section.
Mathematically this problem is identical to the motion of a 
wave-packet in a crystal with noise (temporal randomness).
It also bears formal resemblance to the directed polymer model,
an important problem in statistical mechanics. Hence it is
a problem 
of general interest and has been studied since at least the
1970s from various points of view (see ref \cite{saul} and
refs therein). A considerable amount
is now known. 

For the Saul, Kardar and Read model wave-packet dynamics
can be studied using the mapping to the $q$-model; indeed
the mapping was introduced for this purpose. In this section
we will formulate the problem and summarize known results.
These results reveal that the $q$-model
and the continuum wave model introduced in the last section
behave in qualitatively similar ways.

Consider an electron localized at $n=0$ at $t=0$. This wavepacket
can be propagated to the right using eq (143). As it propagates
it will broaden and its mean position will deflect. It is interesting
to know how the breadth and deflection grow with displacement and to 
analyse the distribution of ``load'' at sufficiently great displacement
that a steady state is reached. 

The root mean square width of the wave-packet grows as the square
root of the displacement. This was derived for the continuum model
in the 1970s \cite{1970s} 
and it is easy to show that the same form is obtained
in the Saul, Kardar and Read model. The root mean square 
deflection grows as the fourth root of the displacement. This
result has been obtained numerically and analytically for both
the Saul, Kardar and Read \cite{saul,yikuo,shapir} and continuum models
\cite{bouchaud,heisenberg}. 

To compare the distribution of load, for the continuum model we
define the load on an edge as $w_n (t) = | \psi_n (t) |^2 $. The
asymptotic distribution of load, $\Pi(w, t \rightarrow \infty)$ 
was obtained by Coppersmith {\em et al.} for the $q$-model \cite{coppersmith}. 
For various
distributions of the fractions, $P(f)$, they found that 
$\Pi(w, t \rightarrow \infty)$ decayed exponentially with
$w$ with a power law prefactor that depended on the distribution
$P(f)$. For the uniform distribution the prefactor was a constant.
The corresponding result for the continuum wave model was obtained
by ref \cite{multifractal} by mapping the problem onto an su(1,1)
quantum ferromagnet. Here too the result for the load distribution
is an exponential with a prefactor linear in $w$. 

\subsection{Field Theory Formulation}

We have emphasized above that the equivalence between the Saul,
Kardar and Read model and the $q$-model is useful only when open
boundary conditions are imposed in the horizontal $t$-direction;
it breaks down for periodic boundary conditions needed to describe
transport in phase-coherent multilayers. The importance of boundary
conditions is also reflected in field theory formulations of these
models. In ref \cite{heisenberg} the continuum model with open 
boundary conditions was mapped onto a Heisenberg ferromagnet.
In contrast, with periodic boundary conditions a mapping to a
supersymmetric analogue of the Heisenberg ferromagnet was obtained
in refs \cite{ilya,zirnbauer}

In this section we derive the supersymmetric spin representation
following the operator methods of ref \cite{heisenberg}. This derivation
highlights the role of boundary conditions, the feature we wish to
emphasize here. It only makes use of operator methods and is in this
sense more elementary than the functional methods of ref \cite{zirnbauer}.
Moreover mappings to supersymmetric spin models have recently been
used fruitfully not only to study the multilayer but also to
provide non-perturbative insights into various other problems
of electron localization \cite{susy,balents,antiparticle}. 
It is hoped that the present derivation,
with its emphasis on boundary conditions\footnote{For the effect 
of boundary conditions on the supersymmetry mapping for models such
as the Chalker model of the quantum Hall transition see refs 
\cite{balents} and \cite{antiparticle} where the random hopping
model in one-dimension is analysed with periodic and open
(scattering) boundary conditions respectively.} and its use of
operator methods will prove of interest in this broader context
also.

\subsubsection{Fermion Representation}

We wish to evaluate $G(n,t;n',t')$, the Green's function
for the continuum model governed by the Schr\"{o}dinger
equation,
\begin{eqnarray}
- i \frac{\partial}{\partial t} G(n,t; n',t') & = &
m_n(t) G(n+1, t; n',t') 
\nonumber \\
 & & + m^{*}_{n-1} (t) G(n-1,t; n',t') 
\nonumber \\
& & - i \delta(t-t') \delta_{nn'},
\end{eqnarray}
and subject to the periodic boundary condition
\begin{equation}
G(n,t+T;n',t') = G(n,t;n',t').
\end{equation}
Here $T$ is the period in the $t$-direction.
In ref \cite{heisenberg}, the Green's function was calculated
subject to the chiral boundary condition, $G(n,t;n',t') = 0$ for
$t < t'$, leading to a simpler field theory formulation.

The key idea is to reinterpret the co-ordinate $t$ as time.
Eq (144) then describes a particle on a one-dimensional lattice
with noise. In second quantised notation the (time-dependent) Hamiltonian
that governs the motion of this fictitious particle is
\begin{equation}
H^R_F (t) = \sum_n \left[ m_n (t) c_n^{R \dagger} c_{n+1}^{R}
+ m_{n-1}^{*} (t) c_{n}^{R \dagger} c_{n-1}^{R} \right].
\end{equation}
Here $c_n^{R \dagger} $ creates a Fermion at site n; $c_n^{R}$ annhilates
it. The reasons for the superscript on the Fermion Hamiltonian and on the
creation and annhilation operators will become apparent shortly.

The $S$-matrix for this model obeys
\begin{equation}
- i \frac{\partial}{\partial t} S^R_F(t) = H_{F}^R (t) S^R_F(t)
\end{equation}
subject to $S^R_F(t \rightarrow 0) = 1$. From $S_F^{R-1} S_F^R = 1$ it is easy
to verify the useful result
\begin{equation}
i \frac{\partial}{\partial t} S_F^{R-1} = S_F^{R-1}(t) H_F^R(t).
\end{equation}
We define 
\begin{equation}
c_n^{R} (t) = S^{R-1}_F (t) c_n S^R_F(t) 
\end{equation}
and similarly for $c_n^{R \dagger}(t)$. 

Now by analogy with finite temperature field theory  \cite{fetter} we write the
Green's function
\begin{eqnarray}
G(n,t;n',t') & = & {\rm Tr} \hspace{1mm} [ S^R_F(T) c_n^{R}(t) 
c_{n'}^{R \dagger} (t') ]/ Z_F^R(T)
\nonumber \\
& & {\rm for} \hspace{2mm} t > t'
\nonumber \\
& = & - {\rm Tr} \hspace{1mm} [ S^R_F(T) c_{n'}^{R \dagger} (t') 
c_n^{R}(t) ]/ Z_F^R(T)
\nonumber \\
& & {\rm for} \hspace{2mm} t < t';
\nonumber \\
Z_F^R(T) & = & {\rm Tr} \hspace{1mm} [S^R_F(T)].
\end{eqnarray}
$Z_F^R(T)$ is analogous to the partition function in finite
temperature field theory.
It is easy to verify that $G$ obeys the differential eq (144)
by making use of eqs (147) and (148) and the commutation
relation
\begin{equation}
[ H_F^R (t), c_n^{R} ] = - m_n(t) c^R_{n+1} - m_{n-1}^{*} (t)
c^R_{n-1}.
\end{equation}

However
\begin{eqnarray}
G(n,T;n',t') & = & {\rm Tr} \hspace{1mm}
[ S^R_F(T) S_F^{R-1}(T) c_n^R S^R_F(T) c_{n'}^{R \dagger} (t') ]/ Z_F^R(T)
\nonumber \\
 & = & {\rm Tr} [ S^R_F(T) c_{n'}^{R \dagger} (t') c_n^R ]/ Z_F^R(T)
\nonumber \\
& = & - G(n,0; n', t').
\end{eqnarray}
Thus $G$ obeys antiperiodic rather than periodic boundary conditions.
This problem is fixed by adding a term to the Hamiltonian
\begin{equation}
H^R_F(t) \rightarrow H^R_F (t) + \frac{\pi}{T} \sum_n c_n^{R \dagger} c_n^R.
\end{equation}
Alternatively we may replace ${\rm Tr} \rightarrow {\rm STr}$ in
eq (150). By STr we mean the trace of an operator over all states with
an even number of fermions minus the trace over states with an odd
number of fermions.

We also need an expression for the complex conjugate of the Green's function
since our ultimate purpose is to calculate the disorder average of 
$|G(n,t;n',t')|^2$, the diffuson propagator. To this end we complex
conjugate eq (144) to obtain the differential equation obeyed by $G^{*}$.
Comparison to eq (144) reveals that we should consider $A$ fermions
governed by the Hamiltonian
\begin{equation}
H_F^{A}(t) = - \sum_n \left[ m_n(t) c_{n+1}^{A \dagger} c_n^A + m_n^{*}(t) 
c_n^{A \dagger} c_n^A \right].
\end{equation}
$G^{*}(n,t;n',t')$ is then given by the right hand side of eq (150)
if we replace $R \rightarrow A$ and ${\rm Tr} \rightarrow {\rm STr}$. 

As might be expected the Hamiltonian for the $A$ fermions is related
to that for the $R$ fermions via a particle hole transformation.
This symmetry between the $R$ fermions and the $A$ holes leads to
an su(2) symmetry in the fermion sector of the complete field theory
formulation that we obtain below (eq 171). It is also at the root
of the supersymmetry of the field theory formulation.

In summary the Green's function with periodic boundary conditions may be 
generated from the second quantized Hamiltonian, $H_F^R(t)$ [eq (153) 
and (146)] using the definition eq (150). The complex conjugate of
the Green's function may be obtained similarly using the Hamiltonian
$H_F^A(t)$ (eq 154). Eq (150) and its A fermion analogue provide exact
formal expressions for the Green's function for a particular realization
of the random tunneling $m_n(t)$. These expressions are not particularly
convenient to average since $m_n(t)$ appears in both numerator and
denominator.

\subsubsection{Boson Representation}

Alternatively we could interpret eq (144) as a time dependent
Schr\"{o}dinger equation for bosonic particles on a one dimensional
lattice. The corresponding ``time''-dependent bosonic Hamiltonian
in second quantized notation is 
\begin{equation}
H^R_B(t) = \sum_n \left[ m_n(t) b_n^{R \dagger} b_{n+1}^{R} +
m_{n-1}^{*}(t) b_n^{R \dagger} b_{n-1}^{R} \right].
\end{equation}
Here $b_n^{R \dagger}$ creates an $R$ boson at site $n$; $b_n^{R}$ annhilates
it. 

The Green's function is now defined as
\begin{eqnarray}
G(n,t; n',t') & = & {\rm Tr} \hspace{1mm} [ S^R_B(T) 
b_n^{R}(t) b_{n'}^{R \dagger} (t') ]/ Z_B^R(T)
\nonumber \\
 & & {\rm for} \hspace{2mm} t > t'
\nonumber \\
& = & {\rm Tr} \hspace{1mm} [ S^R_B(T) b_{n'}^{R \dagger} (t')
b_n^{R}(t)  ]/ Z_B^R(T)
\nonumber \\
 & & {\rm for} \hspace{2mm} t < t';
\nonumber \\
Z_B^R(T) & = & {\rm Tr} \hspace{1mm} S^R_B(T).
\end{eqnarray}
Here $S^R_B$ is the bosonic S-matrix and $Z^R_B(T)$ is
the bosonic analogue of the partition function.

For greater rigour we must regulate the traces to ensure
convergence but for brevity we do not discuss this explicitly
here.

The complex conjugate of the Green's function is generated similarly
if instead of the R bosons we consider A bosons governed by
\begin{equation}
H_B^A(t) = - \sum_n [ m_n(t) b_{n+1}^{A \dagger} b_n^A 
+ m_n^{*}(t) b_n^{A \dagger} b_{n+1}^A ].
\end{equation}

The main result of this subsubsection is eq (156). It provides a formal
bosonic expression for the exact Green's function for a particular realization
of random tunneling, $m_n(t)$. A similar expression for $G^{*}$ may be
obtained by working with the Hamiltonian eq (157). Like their fermionic 
counterparts these bosonic expressions are not particularly well suited
for averaging over disorder.

\subsubsection{Supersymmetry}

We now develop an expression for the diffuson suitable for averaging
over disorder. In Appendix E it is shown that 
\begin{equation}
Z_F^R(T) Z_B^R(T) = 1.
\end{equation}
Thus we consider a model that includes both A and R fermions and
bosons governed by the Hamiltonian
\begin{eqnarray}
H_{{\rm SUSY}} (t) & = & H_F^R(t) + H_F^A(t) + H_B^R(t) + H_B^A(t)
\nonumber \\
 & = & \sum_n [ m_n (t) A_n + m_n^{*} (t) A_n^{\dagger}].
\end{eqnarray}
Here
\begin{equation}
A_n = c_n^{R \dagger} c_{n+1}^R 
- c_{n+1}^{A \dagger} c_n^A 
+ b_n^{R \dagger} b_{n+1}^R
- b_{n+1}^{A \dagger} b_n^A.
\end{equation}

The corresponding S-matrix obeys
\begin{equation}
- i \frac{\partial}{\partial t} S_{{\rm SUSY}} (t ) = H_{{\rm SUSY}} (t)
S_{{\rm SUSY}} (t)
\end{equation}
subject to $S_{{\rm SUSY}} (t \rightarrow 0) = 1$. A formal solution to
eq (161) is given by
\begin{equation}
S_{{\rm SUSY}} (t) = P \exp \left( i \int_0^t d t_1 H_{{\rm SUSY}} (t_1)
\right).
\end{equation}
Here $P$ is the chronological ordering operator.

Hence the diffuson is given by
\begin{eqnarray}
| G(n,t;n',t') |^2 & = & {\rm STr} \hspace{1mm} [ S_{{\rm SUSY}} (T)
c^{R}_n (t) c^A_n (t) c^{A \dagger}_{n'} (t') c_{n'}^{R \dagger} (t')]
\nonumber \\
 & & {\rm for} \hspace{2mm} t > t'
\nonumber \\
 & = & {\rm STr} \hspace{1mm} [ S_{{\rm SUSY}} (T)
c^{A \dagger}_{n'} (t') c_{n'}^{R \dagger} (t')
c^{R}_n (t) c^A_n (t) ] 
\nonumber \\
 & & {\rm for} \hspace{2mm} t < t'
\end{eqnarray}
The content of eq (163) is that to calculate the diffuson we must
create or annhilate a pair of R and A fermions (depending on the time
order). Then we must propagate this state in accordance with $H_{{\rm SUSY}}$
and perform an S-matrix weighted trace. The Hamiltonian $H_{{\rm SUSY}}$
is non-interacting but it is random and time dependent.

Eq (163) is an exact formal expression for the diffuson. Note the lack
of a denominator, eliminated by virtue of eq (158). This feature allows
us to perform the average over disorder easily. For example,
\begin{equation}
\langle S_{{\rm SUSY}} (t) \rangle = \exp [ - {\cal H}_{{\rm SUSY}} t ]
\end{equation}
with 
\begin{equation}
{\cal H}_{{\rm SUSY}} = \frac{D}{2} \sum_n ( A_n^{\dagger} A_n
+ A_n A_n^{\dagger} ).
\end{equation}
Here we have assumed that the tunneling $m_n(t)$ is a Gaussian
white noise process with zero mean and variance
\begin{equation}
\langle m_n^{(\alpha)} (t) m_{n'}^{(\beta)} (t') \rangle
= D \delta (t - t') \delta_{nn'} \delta_{\alpha \beta}.
\end{equation}
Here $m_n^{(1)}(t) =$ real part of $m_n(t)$; $m_n^{(2)} (t) = $
imaginary part of $m_n(t)$.

Recall that for a single Gaussian random variable $y$, the phase
average $ \langle e^{iy} \rangle = e^{-\langle y^2 \rangle/2}$.
Eq (164) is analogous to this result but with the added complications
that $S_{{\rm SUSY}}$ is an ordered exponential, not a simple exponential,
and the average is over a random process rather than a single random
variable. To derive eq (165) it is simplest to expand the time ordered
exponential (eq 162) and average term by term.

Proceeding in this manner we obtain an expression for the average
diffuson
\begin{eqnarray}
\langle | G(n,t;n',t') |^2 \rangle & = & {\rm STr} \hspace{1mm}
\{ \exp [ - {\cal H}_{{\rm SUSY}}( T - t + t' ) ]
c_n^{R} c_n^A 
\nonumber \\
& & \times
\exp[ - {\cal H}_{{\rm SUSY}} (t - t') c_n^{A \dagger} c_n^{R
\dagger} \}
\nonumber \\
 & & {\rm for} \hspace{2mm} t > t'.
\end{eqnarray}  
A similar expression may be written for the case $t < t'$. The content
of eq (167) is that to calculate the {\em average} diffuson we must
create (or for the other time order, annhilate) a pair of R and A
fermions and propagate the resulting state according to the effective
Hamiltonian ${\cal H}_{{\rm SUSY}}$. In contrast to $H_{{\rm SUSY}}$
the effective Hamiltonian is not time dependent or random but it is
interacting. 

This completes our formulation of the continuum directed wave model of
section VI A as a superspin field theory. The main results are the 
superspin Hamiltonian (eq 165) and eq (167) which shows how interesting
correlation functions are calculated in the superspin formulation.
The usefulness of this formulation depends on the extent to which the
superspin model can be analysed. 

In the remainder of this section we discuss the form and symmetry
of the superspin Hamiltonian (eq 165). To this end it is helpful to
introduce special notation for the boson and fermion bilinears of which
${\cal H}_{{\rm SUSY}}$ is composed. We denote the fermion bilinears
\begin{eqnarray}
J_+ = c^{R \dagger} c^{A \dagger} = J_x + i J_y, & &
\nonumber \\
J_- = c^A c^R = J_x - i J_y, & &
\nonumber \\
J_z = \frac{1}{2} ( c^{R \dagger} c^R + c^{A \dagger} c^A - 1 ), & &
\nonumber \\
J = \frac{1}{2} ( c^{R \dagger} c^R - c^{A \dagger} c^A + 1 ); & &
\end{eqnarray}
the boson bilinears,
\begin{eqnarray}
K_+ = b^{R \dagger} b^{A \dagger} = K_x + i K_y, & &
\nonumber \\
K_- = b^A b^R = K_x - i K_y, & &
\nonumber \\
K_z = \frac{1}{2} ( b^{R \dagger} b^R + b^{A \dagger} b^A + 1), & &
\nonumber \\
K = \frac{1}{2} ( b^{R \dagger} b^R - b^{A \dagger} b^A - 1); & &
\end{eqnarray}
and the mixed bilinears, 
\begin{eqnarray}
M_1 = b^{R \dagger} c^R, \hspace{2mm} M_2 = b^{A \dagger} c^A, & &
\nonumber \\
L_1 = b^{A \dagger} c^{R \dagger}, \hspace{2mm} L_2 = b^{R \dagger} 
c^{A \dagger}. & &
\end{eqnarray}
In eqs (168), (169) and (170) the site indices have been suppressed for
brevity.
In terms of these bilinears we may write
\begin{eqnarray}
{\cal H}_{{\rm SUSY}} & = & - 2 D \sum_n \left( \vec{J}_{n+1}.\vec{J}_n 
+ J_{n+1} J_{n} - J_n \right)
\nonumber \\
& & + 2 D \sum_n \left( \vec{K}_{n+1}.\vec{K}_n + K_{n+1} K_n + K_n \right)
\nonumber \\
& & + D \sum_n \left( M^{(1) \dagger}_{n+1} M^{(1)}_n 
+ M^{(2) \dagger}_{n+1} M^{(2)}_n + {\rm hc} \right)
\nonumber \\
& & + D \sum_n \left( L^{(1) \dagger}_{n+1} L^{(1)}_{n}
+ L^{(2) \dagger}_{n+1} L^{(2)}_n + {\rm hc} \right).
\end{eqnarray}
Here hc denotes Hermitian conjugate and $ \vec{K}_{n+1}.\vec{K}_n =
K_{n+1}^z K_n^z - K_{n+1}^x K_n^x - K_{n+1}^y K_n^y $ .

It is instructive to study the commutation relations for bilinears at
the same site $n$ (bilinears at different sites simply commute or anticommute).
It is easy to verify that $J_+, J_-$ and $J_z$ satisfy angular momentum
or su(2) commutation relations and $J$ commutes with the other three. 
Similarly $K_+, K_-$ and $K_z$ satisfy the su(1,1) or hyperbolic angular
momentum algebra---essentially the angular momentum algebra but with
a sign change for the $K_+, K_-$ commutator \cite{mattis}. 
$K$ commutes with the other
three. The anticommutators of $L_i, L_i^{\dagger}, M_i$ and $M_i^{\dagger}$
are linear combinations of the $K$'s and $J$'s. The commutators of the
$J$'s or $K$'s with the $L$'s or $M$'s are linear combinations of the 
$L$'s and $M$'s. Hence these bilinears constitute a superalgebra. The
$J$'s and $K$'s are commuting elements of the superalgebra; the $L$'s
and $M$'s, anticommuting elements. The superalgebra is called u(1,1|2).
It includes the Lie algebras su(2) and su(1,1) as subalgebras.

Further insight into the superalgebra is obtained by considering the 
Hilbert space at a single site. This is a direct product of the four
dimensional fermion space and the infinite dimensional two-boson space.
The fermion space may be decomposed into irreducible representations
of the su(2) algebra. The fermion vacuum and the state with both
R and A fermions present 
constitute a doublet or spin 1/2 representation; the two states with one
fermion present are singlets. The boson space
similarly decomposes into an infinity of infinite dimensional irreducible
representations of the su(1,1) algebra\footnote{Let $|n+m,n \rangle$
denote a state with $(n+m)$ R-bosons and $n$ A-bosons on the site.
The infinite dimensional subspace with $m$ a fixed integer and $n = 0,
1, 2, \ldots$, for $m \ge 0$, or $n = -m, -m+1, =m+2, \ldots$, for
$m < 0$, is invariant under the four $K$ operators. These subspaces
corresponding to different values of $m$ constitute the irreducible
representations of the su(1,1) algebra.}.
The single site Hilbert space thus decomposes rather simply into irreducible
representations of the direct sum of the su(2) and su(1,1) algebra.
These subspaces do not constitute a representation of the whole superalgebra.
The anticommuting elements mix different irreducible representations of
su(2) and su(1,1). In particular they mix representations with different
spins---a celebrated feature of supersymmetry. It is not difficult to
decompose the single site Hilbert space into blocks irreducible under the
superalgebra; however this would carry us too far afield. More details
on the superalgebra are given in ref \cite{zirnbauer} and refs therein.

Finally we define
\begin{equation}
{\cal J}_{{\rm tot}} = \sum_n {\cal J}_n.
\end{equation}
Here ${\cal J}$ denotes any element of the superalgebra such as 
$J_+, K_z, L_1$ etc. After some algebra we find
\begin{equation}
[ {\cal H}_{{\rm SUSY}}, {\cal J}_{{\rm tot}} ] = 0
\end{equation}
revealing the supersymmetry of the field theory formulation.

\section{Summary and Conclusion}

Much of this paper is concerned with the behaviour of
the $q$-model close to the critical point. To probe this
behaviour we imagine that a uniform load is applied to
the top layer. As the load propagates downward fluctuations
develop in the distribution of load. 
Coppersmith {\em et al.} \cite{coppersmith}
studied the entire distribution of load at very great depth where 
it was presumed that a steady state had been reached. In
contrast we study only the variance of the distribution of
load but we analyse its evolution with depth. Our purpose is
to study this evolution for all distributions of the fractions, 
$P(f)$, particularly
those close to the singular distribution (the critical point).

In section II we consider the $q$-model in 1+1 dimensions
without injection (the weight of the beads is neglected). 
In this case the average load does not vary with
depth since the total load is the same in every layer; it is
merely redistributed by the q-model dynamics. For the growth
of the variance, by analogy to critical phenomena, we make the
following hypotheses: For all distributions $P(f)$ except the
singular distribution we posit that the variance will saturate
at sufficient depth. Both the saturation depth and the
saturated variance are expected to diverge as the distribution
approaches the singular distribution. We introduce $\delta$, a
measure of the distance of a distribution $P(f)$ from the
singular distribution, and conjecture that the saturation depth
$\xi_{{\rm corr}}$ will diverge as $1/\delta^{\varphi}$; the saturated
variance, as $1/\delta^{\theta}$. More specifically, we expect that
close to the critical point the variance will have a scaling form,
eq (7). For the singular distribution we expect that the variance
will grow indefinitely as a power of the depth. Close to the critical
point and at depths shallow compared to the saturation depth the
variance should grow as it would right at the critical point. From this
and from eq (7) we deduce a relationship between the critical exponents
$\theta$ and $\varphi$ and the exponent that describes the growth of
the variance right at the critical point; namely we expect that 
at the critical point the variance will grow as $t^{\theta/\varphi}$. In
sections IIB and IIC we derive an exact formula for the variance as
a function of depth (eq 46) and study its scaling limit ($t \gg 1$,
$\delta \rightarrow 0$ but with $t \delta^{\varphi}$ arbitrary). These
calculations bear out all the expectations enumerated above, provide
the precise form of the scaling function [eq (50) and Fig 4] and
yield the exact exponents (eq 49).

In section III we characterise the critical point more fully
by analysing the evolution with depth of the entire distribution
of load right at the critical point in 1+1 dimensions. In the 
absence of injection the critical point is a simple model of
random walkers that coalesce upon contact; hence it is quite 
straightforward to derive these results. We present them because 
they illuminate the results of the previous section.
At large depth it is found that the distribution of load
consists of a large spike at zero load together with a
smooth part [eq (73) and (76)]. It is overwhelmingly probable that the load
on a bead is zero; most of the weight of the distribution
is in the spike. The smooth part follows the anticipated scaling
form (eq 53). Its width grows as the square root of the depth,
consistent with the exponent found in section II to describe the
growth of the variance of load at the critical point.

In section IV the effect of injection is included. For simplicity
we consider only 1+1 dimensions. We assume that
the weights of the beads are independent and identically distributed 
random variables. The behaviour of the mean load is still not
very interesting. It grows linearly with depth (eq 77). Close to
the critical point we conjecture that the variance will have the
form eq (78).  We are able to deduce all the exponents
in eq (78) and to obtain some limiting behaviours of the scaling
function through simple (non-rigourous) arguments. These
conjectures are all verified by the exact calculation of sections
IV A and IV B which provides the precise form of the scaling
function [eqs (99), (102) and (103)] and yields all the exponents
[eqs (80), (81) and (84)]. We find that beyond a crossover depth
the variance
(normalized by the squared mean) saturates. The saturation value 
and the crossover depth both diverge
as the critical point is approached. At depths less than the crossover
depth the variance grows as it would right at the critical point
(eq 106). 
The behaviour at the critical point has many crossovers if the
weight of the beads is small compared to the applied load. In this
case at first the variance grows as the square root of the depth
as it was found to do in section II in the absence of injection.
At greater depths there are crossovers to growth as $t$
and $t^{5/2}$, as first 
the effects of large rare fluctuations in the weight of a
bead and  then mean injection assert themselves. Ultimately at the critical
point the
variance grows with the 5/2 exponent but the depth at which
this behaviour sets in can be very great if the mean injection
is small. This depth diverges as $\langle I \rangle^{-4/3}$. The
crossover exponent 4/3, deduced by simple arguments and then
via exact calculation in section IV, agrees with the value previously
obtained by a different method by Majumdar and Sire \cite{clement}.
In their work Majumdar and Sire only study the behaviour right
at the critical point. However at this point they calculate
the dynamics of the entire distribution of load whereas we
study only the variance.
 
In section V we turn to the $q$-model in D+1 dimensions. For
simplicity we neglect injection in this section.
We find that right at the critical point the variance grows 
as a power of depth in all dimensions except two (eq 137). The power is given
by $D/2$ for $D < 2$. For all dimensions above two the growth is
linear. This shows that $D=2$ is the upper critical dimension
for this problem. For $D=2$ we find a linear growth of the
variance tempered by a log factor as might be expected at
the critical dimension. 

An intriguing feature of the critical behaviour we obtain is that
it is exhibited at all. For ordinary continuous phase transitions
the renormalization group provides a framework to understand
the critical behaviour. We are not aware of any such framework
for the $q$-model.

Random critical points are notoriously difficult to analyse
in general.
The feature that allows us to analyse the $q$-model is
that the two point load correlation function [defined by eqs (8) and
(114)] evolves with depth according to a simple linear equation.
In section II we analyse the evolution by expanding in 
the eigenvectors of an appropriate linear operator.
There are some subtleties posed by the non-Hermiticity
of the linear operator, making it necessary to prove
that its eigenvectors are complete (further complicated by
the infinite dimensionality of the vector space). Nonetheless 
we like this approach because it parallels transfer matrix
methods used for equilibrium critical phenomena. We find that the
large depth scaling behaviour is controlled by the low energy
long wavelength
eigenfunctions of the non-Hermitian ``Hamiltonian''. Another
virtue of this approach is that with about the same effort
it yields both the variance and the correlation functions.
However we have left analysis of the correlation functions
open for later work. Here we focus entirely on the variance
of the load. In section V we analyse the variance using another
technique based on transform methods.

Our analysis, neglecting injection, 
confirms that the q-model has essentially no
horizontal correlations in the steady state for any distribution
except the singular. This agrees with experiments on bead packs.
The bulk of our results however are concerned with the q-model
close to the critical point. Bead pack experiments such as those
of ref \cite{jaeger} appear to be far from the critical point.
We estimate $\delta \approx 0.5$ for this experiment. It is
not obvious how to tune the parameter for bead packs to 
access the critical behaviour we analyse here.
Claudin {\em et al.} have also studied the horizontal steady state 
correlations of the $q$-model without injection away from the
critical point \cite{claudin}. They employ
a continuum limit and arrive at conclusions similar to eq (19)
in section IIB. The main focus of their work however is to explore 
a tensor model of stress propagation in granular matter, intended to
supplant the $q$-model.

Interpreted in terms of river networks our results show that
allowing a small amount of river splitting in a Scheidegger
network introduces a length scale in the vertical direction.
On sufficiently long length scales such a network is not
scale invariant. This resembles the finding
of Narayan and Fisher \cite{onuttom}. In their model too there was a parameter
that controlled river splitting. Their networks were not
scale invariant unless river splitting was tuned to zero.
However their model appears to be in a different universality
class as its vertical correlation exponent is different from
the value $\varphi = 2$ we obtain here in section II.
Presumably the difference is because their rule for
stream splitting was non-local and depended on the 
entire history of the network upstream from the split.

Taken together with the model of Narayan and Fisher it appears
that river splitting is a perturbation that spoils the scale 
invariance of river networks. It is therefore interesting to ask
whether such networks exist in Nature. River deltas are one
possibility. Traced backwards they may constitute networks
of merging streams that occasionally split. Even for river basins
it might be interesting to examine the extent to which streams
split. In this context it is worth noting that some of the data
against which river scaling laws are tested are based not on actual
maps of the river network but on networks that are indirectly
inferred according to certain rules
from digital elevation data obtained from satellite images. 
The rules by which the network is inferred from the elevation maps 
exclude the possibility of splitting \cite{riverbook}.

In summary the $q$-model is rich in applications and behaviour
and yet analytically tractable by elementary means a combination of
circumstances that invites
further exploration. Among the many problems that remain open we
conclude by mentioning two:
For the $q$-model beyond the saturation depth there
is no correlation in the horizontal direction but in the
vertical direction there are very strong and long ranged correlations
\cite{nagel}. We have not obtained
the precise form of these vertical correlations for the $q$-model
either in steady state or at the critical point. It would be very
interesting (and straightforward) to obtain these forms and the
crossover between them. Second it would be interesting to obtain the
dynamics of the entire distribution of load near the critical
point. We have not attempted to do this
except right at the critical point.

A natural scaling hypothesis is that the full distribution of load,
neglecting injection, will be of the form
\begin{equation}
\Pi(w,t,\delta) = w^{-\tau} {\cal Q}( w \delta^{1/\sigma}, t \delta^{\nu z}).
\end{equation}
The exponents in eq (174) are $\tau = 2, \nu z = 2$ and
$\sigma = 1$. Their values are fixed by our result for the variance
away from the critical point derived in section II and the result 
for the entire distribution at the critical point $\delta = 0$
derived in section III. We also know that for $x \rightarrow 0$
and $y \rightarrow 0$, the presently unknown function
${\cal Q}$ has the asymptotic behaviour
\begin{equation}
{\cal Q}(x,y) \approx \frac{4}{\sqrt{\pi}} \frac{x}{y^{3/2}} e^{- x^2/y }
\end{equation}
to be consistent with the critical point distribution (eq 73) derived
in section III.

In the second part of this paper we turn to chiral wave models
that are believed to describe the surface electronic states of a
quantum Hall multilayer. In section VI A we discuss circumstances
under which the quantum network model of Saul, Kardar and Read
(described in the introduction) is equivalent to the $q$-model.
In section VI B we compare known results about the behaviour
of the $q$-model to a continuum chiral wave model that cannot
be mapped onto a $q$-model under any circumstance. The two 
are found to behave in qualitatively similar ways.

A circumstance under which the mapping to the $q$-model is not
useful is when periodic boundary conditions must be imposed in
the chiral direction. Physically this is because of the interference
of electron paths that wind around the quantum Hall multilayer.
Such long range interference cannot be captured by the classical
$q$-model. In this phase-coherent or mesoscopic regime, the
chiral wave model has been studied via a mapping to a supersymmetric
spin model \cite{ilya,zirnbauer}. 
In section VI C we derive this mapping in a way that
emphasizes boundary conditions. Our derivation makes use of
operator methods and is hence more elementary than the derivation
of ref \cite{zirnbauer} that makes use of mixed functional integrals
over Grassman and bosonic variables. We do not attempt further analysis
of the superspin model here; the interested reader should
consult papers on multilayer transport, particularly refs
\cite{cho} and \cite{ilya} that provide a nice
overview of the early work on this problem.

Mappings to superspin models have been useful not only in
the study of the quantum Hall multilayer but have also
recently lead to new non-perturbative results and insights
into other important problems of electron localization 
\cite{susy,balents,antiparticle}. Hence
it is hoped that our derivation, with its emphasis on boundary
conditions and use of elementary operator methods, will be of
interest in this general context.

{\em Note added:} While writing this paper we learnt of an {\em e}-print
by Rajesh and Majumdar on spatio-temporal correlations in
the Takayasu model and the $q$-model \cite{rajesh}. These
authors derive many interesting results complementary to ours.
In this paper we concentrate on the behaviour close to the 
critical point. For the $q$-model Rajesh and Majumdar 
concentrate on length scales long compared to our vertical
correlation length, $\xi_{{\rm corr}}$; the crossovers
and scaling functions that we study are transients that are invisible in their 
asymptotic formulae. On the other hand they have derived both
vertical and horizontal load correlation functions; this paper
is limited (in practice but not in principle) to the study
of the variance of load. Among their interesting findings
are (i) They find power law correlations in the vertical direction
both at the critical point and away from it addressing in part a
question raised above. (ii) They emphasize the interesting structure
of the horizontal correlation function, including injection, at 
great depth.

Although the goals are a bit different, 
there are points of intersection between the two
papers with regard to technique. Rajesh and Majumdar 
too exploit the linearity of the relation that describes
the evolution of the correlations with depth and solve it using
the method of section V. An overlapping result is a formula for
the variance at the critical point in 1+1 dimensions including injection.
At the large depths studied by Rajesh and Majumdar the last term in
our eq (106) should dominate. Rajesh and Majumdar obtain the same
exponent 5/2 and the same numerical prefactor $16/15\sqrt{\pi}$ providing
a nice check on both calculations.

It is a pleasure to acknowledge stimulating discussions with
Sue Coppersmith and Onuttom Narayan. We thank Onuttom Narayan in
particular for patient explanation of refs \cite{coppersmith},\cite{narayan} 
and \cite{onuttom}, for 
encouraging us to study the critical point and
for explaining to us the significance of river splitting.
This work was supported in part by NSF Grant
DMR 98-04983 and by the Alfred P. Sloan Foundation. HM acknowledges
the hospitality of the Aspen Center for Physics where this work 
was completed.

\appendix

\section{Proof of Completeness}

First let us recall the principles of biorthogonal expansion (see,
for example, ref \cite{morse}, p884).
We discuss the simplest case of a finite $N \times N$
dimensional non-Hermitian matrix $H_{mn}$. Consider its
eigenvectors
\begin{equation}
\sum_n H_{mn} \phi_{n}^{\lambda} = \lambda \phi_{m}^{\lambda}.
\end{equation}
In context of biorthogonal expansion these eigenvectors are
called the right eigenvectors. 
For simplicity we will assume that the right eigenvalues are 
non-degenerate in this case. The bad news regarding the right
eigenvectors is: (i) $\lambda$ may be complex. (ii) There is no
guarantee that there are $N$ eigenvectors (needed to span the
vector space). (iii) Eigenvectors corresponding to different 
eigenvalues are not necessarily orthogonal. 

Now consider the left eigenvectors, defined as the eigenvectors
of $H^{\dagger}$. (i) If $\lambda$ is a right eigenvalue then
$\lambda^{*}$ is a left eigenvalue (Proof: The coefficients for
the characteristic polynomials of $H$ and $H^{\dagger}$ are
complex conjugates of one another). (ii) There are as many left
eigenvectors as right. (iii) Left eigenvectors are orthogonal to
right eigenvectors.

The last point merits elaboration. Let $\psi_n^{\lambda}$ denote
the left eigenvector with left eigenvalue $\lambda^{*}$. Thus
\begin{equation}
\sum_n H^{\dagger}_{mn} \psi_n^{\lambda} = \lambda^{*} 
\psi_{m}^{\lambda}.
\end{equation}
According to (iii) above
\begin{equation}
\sum_n ( \psi_n^{\lambda} )^{*} \phi_n^{\lambda'} = 
\delta_{\lambda \lambda'}.
\end{equation}
Eq (A3) is the {\em biorthogonality} relation. It may be
proved by noting
\begin{eqnarray}
\sum_{mn} ( \psi_{n}^{\lambda} )^{*} H_{nm} \phi_{m}^{\lambda'} 
& = & \lambda \sum_{n} ( \psi_{n}^{\lambda} )^{*} \phi_{n}^{\lambda'}
\nonumber \\
 & = & \lambda' \sum_{n} ( \psi_{n}^{\lambda} )^{*} \phi_{n}^{\lambda'}
\end{eqnarray}
whence $ \sum_{n} ( \psi_{n}^{\lambda} )^{*} \phi_{n}^{\lambda'} = 0$
for $\lambda \neq \lambda'$. 

In general there is no guarantee of completeness, but in this case
assume that $N$ eigenvectors have been found. Then we can prove
the {\em completeness} relation
\begin{equation}
\sum_{\lambda} ( \psi_{m}^{\lambda} )^{*} \phi_{n}^{\lambda}
= \delta_{mn}.
\end{equation}
The proof follows from the observation that if there are $N$ 
eigenvectors, any vector $a_n$ may be expanded as 
\begin{equation}
a_n = \sum_{\lambda} a_{\lambda} \phi_n^{\lambda}.
\end{equation}
Completeness then follows from biorthogonality, eq (A3).

The problem in section IIB presents some complications not
present in the pedagogical discussion above. Among them are
degeneracy, an infinite dimensional vector space and a continuous
spectrum. Nonetheless the broad strategy is the same. In section IIB
we found left and right eigenvectors and we conjectured biorthogonality
and completeness relations. To justify the analysis of section IIB
we must prove the completeness relation. That is the purpose of this
appendix. Note that we cannot simply assume completeness is true---because
the matrix $H$ is non-Hermitian there are no theorems to guarantee
it. Nor can we prove completeness by counting eigenvectors as in the
finite dimensional discussion above.

The proof of completeness is remarkably simple and direct.
We substitute the exact expressions for $\psi^{(\pm) k}_m$
and $\phi^{(\pm) k}_n$ that we have derived, eqs (28), (31),
(33) and (34), on the right hand side of eq (36) and verify
the completeness relation by explicit evaluation of the integral.
There are nine cases to consider corresponding to $n=0, n>0, n<0$
and $m=0, m<0, m>0$.

For illustration we analyse the case of $n=0, m=0$. We must evaluate
\begin{equation}
\frac{2}{\pi} \int_{0}^{\pi} d k {\cal A}^{*}(k) A(k)
\end{equation}
where ${\cal A}(k)$ and $A(k)$ are as given in eqs (30) and (35).
Since the integrand is symmetric in $k$ we extend the range of integration
from $-\pi$ to $\pi$ and substitute $z \rightarrow e^{ik}$ to obtain
a contour integral about the unit circle
\begin{equation}
\oint \frac{d z}{2 \pi i} \frac{1}{z} \frac{ (1 - \epsilon) 
(z + 1)^2 }{ \epsilon^2 (z - 1)^2 - (1 - \epsilon)^2 (z + 1)^2 }.
\end{equation}
Evaluation via Cauchy's theorem reveals that the integral equals
one as required for completeness. 

The remaining eight cases also succumb to this method of analysis.

\section{Inverse z-transform}

Consider the series $f(t)$, $t = 0,1,2, \ldots$ Its $z$-transform
is defined as
\begin{equation}
f(z) = \sum_{t=0}^{\infty} f(t) z^t.
\end{equation}
Some $z$-transforms can be inverted by inspection. For example
the inverse transform of $ (1 - \alpha z)^{-1}$ is evidently
\begin{equation}
(1 - \alpha z)^{-1} \rightarrow f(t) = \alpha^t.
\end{equation}
In other cases the inverse transform can be found by performing
the complex integral
\begin{equation}
f(t) = \oint_{C} \frac{d z}{2 \pi i} \frac{f(z)}{z^{t+1}}.
\end{equation}
The contour $C$ must enclose the origin but no singularities of
$f(z)$. 

For illustration let us analyse
\begin{equation}
f(z) = (1 - z)^{-1/2} (1 - \alpha z)^{-1}
\end{equation}
needed to go from eq (43) to (44) in section IIB. Here $\alpha > 1$.
$f(z)$ has a pole at $1/\alpha$ and a branch cut at $z=1$ (see fig 10).
We deform the contour $C$ that encloses the origin to contours $C_1$
and $C_2$ that encircle the pole and pass above and below the branch cut.
Hence obtain
\begin{equation}
f(t) = \sqrt{ \frac{\alpha}{\alpha - 1} } \alpha^t -
\frac{1}{\alpha \pi} \int_{1}^{\infty} d x (x - 1)^{-1/2} 
\frac{1}{x^{t+1}} \left( x - \frac{1}{\alpha} \right)^{-1}.
\end{equation}
The first term is the contribution of the pole; the second, of the
branch cut.

\section{Asymptotics of $\Phi_n(u)$}

The asymptotics of the functions $\Phi_n(u)$ defined by eq (103)
are needed to obtain the asymptotic behaviour of the scaling
functions in sections IIC and IVB. 

The large $u$ behaviour poses no difficulty. Evidently
\begin{equation}
\Phi_n (u) \approx \frac{ \sqrt{\pi} }{2} \frac{1}{\sqrt{u}}
\hspace{2mm} {\rm as} \hspace{2mm} u \rightarrow \infty
\end{equation}
for all $n$. The small $u$ behaviour is a bit more subtle.
Moreover, it turns out that due to cancellations we will need as
many as five or six terms in the small $u$ series for $\Phi_n$
to obtain the leading behaviour of the scaling functions.

For definiteness consider the small $u$ behaviour of
\begin{equation}
\Phi_1 (u) = \int_{0}^{\infty} d s \frac{ e^{-us^2} }{ (1 + s^2) }.
\end{equation}
The leading term is obtained by setting $u=0$,
\begin{equation}
\Phi_1 (0) = \frac{\pi}{2}.
\end{equation}
To obtain the next term it is tempting to expand the integrand
in powers of $u$ but this leads to divergent integrals. The divergence
signals that the asymptotic series is not a simple power series
in $u$.

It turns out the next term goes as $\sqrt{u}$. To show this, and to
efficiently obtain many more terms in the series, consider
\begin{equation}
g(x) = \int_0^{\infty} d s \frac{ e^{-x^2 s^2} }{ (1 + s^2) }.
\end{equation}
We will show that $g(x)$ is regular about $x=0$ and that its
asymptotic behaviour is a simple power series.
To this end we observe that $g(x)$ obeys the first order differential
equation
\begin{equation}
\frac{d}{dx} g - 2 x g(x) + \sqrt{\pi} = 0.
\end{equation}
$x=0$ is a regular point for this equation; hence we attempt
a series solution
\begin{equation}
g(x) = b_0 + b_1 x + b_2 x^2 + \ldots
\end{equation}
We find $b_1 = - \sqrt{\pi}$ and the simple recurrence relation
\begin{equation}
b_n = \frac{2}{n} b_{n-2}.
\end{equation}
Evidently $b_0 = g(0) = \pi/2$.
Hence we obtain the asymptotic series
\begin{equation}
g(x) = \frac{\pi}{2} - \sqrt{\pi} x + \frac{\pi}{2} x^2 
- \frac{2}{3} \sqrt{\pi} x^3 + \frac{\pi}{4} x^4 
- \frac{4}{15} \sqrt{\pi} x^5 + \ldots
\end{equation}
Substituting $x \rightarrow \sqrt{u}$ we conclude
\begin{eqnarray}
\Phi_1(u) & = & \frac{\pi}{2} - \sqrt{\pi} u^{1/2} + \frac{\pi}{2} u
- \frac{2}{3} \sqrt{\pi} u^{3/2} + \frac{\pi}{4} u^2
\nonumber \\
& &
- \frac{4}{15} \sqrt{\pi} u^{5/2} + \ldots
\end{eqnarray}
for small $u$. Similarly
\begin{eqnarray}
\Phi_2 (u) & = & \frac{\pi}{4} - \frac{\pi}{4} u 
+ \frac{2}{3} \sqrt{\pi} u^{3/2} - \frac{3}{8} \pi u^2
+ \frac{8}{15} \sqrt{\pi} u^{5/2} + \ldots
\nonumber \\
\Phi_3 (u) & = & \frac{3 \pi}{16} - \frac{\pi}{16} u
+ \frac{3 \pi}{32} u^2 - \frac{4}{15} \sqrt{\pi} u^{5/2}
+ \ldots
\end{eqnarray}

\section{Lattice Green's Function}

\subsection{Two Dimensions}

Consider the Green's function in two dimensions for the lattice
Schr\"{o}dinger equation discussed in section VA. The real space
Green's function at the origin is given by 
\begin{equation}
{\cal G} (E) = \int_{-\pi}^{\pi} \frac{d k}{2 \pi}
\int_{-\pi}^{\pi} \frac{d p}{2 \pi} \left\{ E -
\frac{1}{4} (1 + \cos p) (1 + \cos k) \right\}^{-1}
\end{equation}
[cf. eq (121) and (128)]. We consider real $E > 1$.
In this appendix we show that 
\begin{equation}
{\cal G}(E) = \frac{2}{\pi E} K \left( \frac{1}{\sqrt{E}} 
\right);
\end{equation}
Here $K$ is a complete elliptic integral of the first kind. From the
well-documented properties of these integrals or by direct analysis
of eq (D8) below it follows that as $E \rightarrow 1^{+}$
\begin{equation}
{\cal G} (E) \approx \frac{1}{\pi} \ln \frac{1}{E - 1}.
\end{equation}
In section V C we are interested in the behaviour of $G(z)$, eq (131),
as the real variable $z \rightarrow 1^{-}$. Comparing eq (131) to (D1)
we see that
\begin{equation}
G(z) = \frac{1}{z} {\cal G} \left( E \rightarrow \frac{1}{z} \right).
\end{equation}
Hence the singularity of $G(z)$ as $z \rightarrow 1$ is
\begin{equation}
G(z) = - \frac{1}{\pi} \ln (1 - z).
\end{equation}
Eq (D5) is the main result of this section of the Appendix.

To demonstrate eq (D2) we regard $p$ as a complex variable
$p \rightarrow x + i y$. The integral over $p$ in eq (D1) may be
regarded as an integral around the contour sketched in Fig 11 since
the two vertical segments cancel by the periodicity of the integrand
and the horizontal segment at infinity makes no contribution because
the integrand vanishes along it. The integrand in eq (D1) has a simple
pole at $p = i y$, where $y$ satisfies
\begin{equation}
\cosh \left( \frac{y}{2} \right) = \frac{ \sqrt{E} }{ \cos (k/2) },
\end{equation}
with residue 
\begin{equation}
\left( i E \sqrt{ 1 - \frac{1}{E} \cos^2 \frac{k}{2} }
\right)^{-1}.
\end{equation}
Hence by Cauchy's theorem
\begin{equation}
{\cal G}(E) = \frac{1}{E} \int_{-\pi}^{\pi} \frac{d k}{2 \pi}
\left( 1 - \frac{1}{E} \cos^2 \frac{k}{2} \right)^{-1/2}.
\end{equation}
Comparing to the definition of the elliptic integral of the first kind
\begin{equation}
K(k) = \int_{0}^{\pi/2} d \theta (1 - k^2 \sin^2 \theta )^{-1/2}
\end{equation}
we obtain eq (D2).

\subsection{Below Two Dimensions}

In this section we analyse the singular behaviour as $z \rightarrow 1$
of $G(z)$ in less than two dimensions. The approximate long wavelength
expression for $G$, eq (133), provides a useful starting point.

To analyse the divergence in $D=1$ we would note that the integrand in (133)
is a sharply peaked Lorentzian. This justifies working to quadratic order
in $S(\vec{p})$ and extending the range of integration (strictly
confined to the Brillouin zone, $-\pi < p < \pi$ in one dimension) to $\pm
\infty$. Result: $G(z) = (1 - z)^{-1/2}$.

To continue this result to non-integral $D$ we use 't Hooft and Veltman's
dimensional regularization trick \cite{stone}. We write
\begin{equation}
G(z) \approx \int_{0}^{\infty} d s \int 
\frac{d \vec{p}}{(2 \pi)^D} \exp - s[ (1 - z) + \vec{p}^2 ];
\end{equation}
extend the range of integration, outside the Brillouin zone and over
all $\vec{p}$-space; and replace
\begin{equation}
\frac{d \vec{p}}{(2 \pi)^D} \rightarrow \frac{ \Omega_D }{(2 \pi)^D}
\int_{0}^{\infty} d p p^{D-1},
\end{equation}
since the integrand in eq (D10) is isotropic in $\vec{p}$. Here
$\Omega_D = 2 \sqrt{\pi}^D/\Gamma(D/2)$ is the total solid angle
in $D$ dimensions (some familiar special cases: $\Omega_1 =2,
\Omega_2 = 2 \pi, \Omega_3 = 4 \pi, \Omega_4 = 2 \pi^2$.) The
result is
\begin{equation}
G(z) = \frac{ \Gamma(1 - D/2) }{\sqrt{\pi}^D} (1 - z)^{D/2 - 1}
\end{equation}
for $D < 2$. 

This analysis breaks down in two dimensions and higher because the 
integrand diverges as $\vec{p} \rightarrow \infty$. The divergence
is an artifact of the quadratic approximation in eq (133) and of
extending the integral outside the Brillouin zone. The spurious
divergence is revealed 
in eq (D12) as a pole in the Gamma function factor as $D \rightarrow
2$. 

\section{Analysis of Partition Function}

The purpose of this appendix is to show that the partition
functions for bosons and fermions cancel. Thus
\begin{equation}
{\rm Tr} \hspace{1mm} [ S_F^R(T) ] 
{\rm Tr} \hspace{1mm} [ S_B^R(T) ]
=1.
\end{equation}
A similar relation holds for the advanced bosons and fermions.
We discuss the retarded case explicitly. For brevity the superscript
$R$ will be omitted. 

We write the fermion S-matrix as
\begin{equation}
S_F(t) = \exp \left( i \frac{ \pi t }{T} \sum_n c_n^{\dagger} c_n
\right) {\cal S}_F (t).
\end{equation}
${\cal S}_F(t)$ is then governed by the Hamiltonian eq (146) without
the extra term included in eq (153).

To make further progress we introduce $e_l^n(t)$, the solution to
the Schr\"{o}dinger eq (144)
\begin{equation}
- i \frac{\partial}{\partial t} e^n_l(t) = m_l(t) e^n_{l+1}(t)
+ m_{l-1}^{*} (t) e^n_{l-1}(t)
\end{equation}
subject to $e^n_l(t \rightarrow 0) = \delta_{nl}$.

The scattering formula
\begin{equation}
c_n {\cal S}_F (t) = \sum_{l} e_n^l(t) {\cal S}_F(t) c_l
\end{equation}
will prove very useful. To derive it, rewrite eq (E4) as
\begin{equation}
{\cal S}_F (t)^{-1} c_n {\cal S}_F (t) =
\sum_l e_n^l (t) c_l
\end{equation}
and regard it as an ansatz with the functions $e^l_n(t)$ 
unspecified. Making use of eqs (147), (148) and (151), the
$t$ derivative of the left hand side is
\begin{eqnarray}
{\cal S}_F(t)^{-1} \{ m_n(t) c_{n+1} +
m_{n-1}^{*}(t) c_{n-1} \} {\cal S}_F(t) & &
\nonumber \\
= \sum_l c_l \{ m_l(t) e^l_{n+1} (t) +
m_{l-1}^{*} (t) e_{n-1}^l (t) \}.
\end{eqnarray}
To obtain the second line we have made use of the ansatz
(E5). Comparing eq (E6) to the $t$ derivative of the right
hand side of eq (E5) we conclude that $e^l_n(t)$ does obey
the Schr\"{o}dinger eq (E3). This completes the proof of
the scattering formula (E4).

Another relation that will prove useful is 
\begin{equation}
{\cal S}_F(t) | 0 \rangle = | 0 \rangle.
\end{equation}
This follows because the Hamiltonian (eq 146) annhilates the
vacuum; ${\cal S}_F$ is the (chronologically ordered) exponential
of the Hamiltonian.

Equipped with these results we write the fermion partition function
as
\begin{eqnarray}
Z_F(T) & = & {\rm Tr} \hspace{1mm} \{ \exp[ i \pi
\sum_n c_n^{\dagger} c_n ] {\cal S}_F(T) \}
\nonumber \\
& = & \langle 0 | {\cal S}_{F}(T) | 0 \rangle
\nonumber \\
& & - \sum_n \langle 0 | c_n {\cal S}_F(T) c_n^{\dagger} 
| 0 \rangle
\nonumber \\
& & + \frac{1}{2} \sum_{n_1, n_2} 
\langle 0 | c_{n_1} c_{n_2} {\cal S}_F(T) c_{n_2}^{\dagger}
c_{n_1}^{\dagger} |0 \rangle
\nonumber \\
& &
- \frac{1}{3!} \sum_{n_1, n_2, n_3} \langle 0 | 
c_{n_1} c_{n_2} c_{n_3} {\cal S}_F(T) 
c_{n_3}^{\dagger} c_{n_2}^{\dagger} c_{n_1} | 0 \rangle
\nonumber \\
& & + \ldots
\end{eqnarray}
The trace is taken over the entire Fock space including states
with different total numbers of fermions. The alternating signs
are due to the factor $ \exp [ i \pi \sum_n c_n^{\dagger} c_n ] $
in the trace. The factorials are because the sums over the site
indices $n_i$ are unrestricted; hence each state gets counted a
multiple number of times. 

We now shift the S-matrix to the left using the scattering formula
(E4), make use of the adjoint of (E7) and calculate the vacuum
expectations of the fermion operators (Wick's theorem). The
result for the second-order term is
\begin{eqnarray}
Z_F(T) & = & \frac{1}{2} \left[ \sum_n e^n_n(T) \right]^2
- \frac{1}{2} \sum_{n_1, n_2} e^{n_2}_{n_1} (T) e^{n_1}_{n_2} (T)
\nonumber \\
& & + {\rm others}.
\end{eqnarray}
Fig 12(b) shows a diagrammatic representation of this term.
Note that the diagram series for the partition function
$Z_F(T)$ contains both connected and unconnected graphs.
By familiar arguments \cite{fetter} we can write
\begin{equation}
Z_F(T) = \exp [ - \Omega (T) ]
\end{equation}
where the ``free energy'' $\Omega(T)$ has the linked diagram
expansion shown in fig 12(c). 
 
We turn now to the boson partition function.

The boson scattering formula
\begin{equation}
b_n S_B(t) = \sum_l e^{l}_n (t) S_B(t) b_l 
\end{equation}
can be proved in the same way as eq (E4). Eq (E7)
remains true when we replace ${\cal S}_F \rightarrow
S_B$

The boson partition function is therefore given by
\begin{eqnarray}
Z_B(T) & = & {\rm Tr} \hspace{1mm} \{ S_B(T) \}
\nonumber \\
& = & \langle 0 | S_B(T) | 0 \rangle 
\nonumber \\
& & + \sum_n \langle 0 | b_n S_B(T)
b_n^{\dagger} | 0 \rangle
\nonumber \\
& & + \frac{1}{2!} \sum_{n_1, n_2} \langle 0 |
b_{n_1} b_{n_2} S_B(T) b_{n_2}^{\dagger}
b_{n_1}^{\dagger} | 0 \rangle
\nonumber \\
& & + \ldots
\end{eqnarray}
This equation resembles eq (E8) but there is an 
extra subtlety in the combinatoric factors.
In the two-boson case for example, for the offdiagonal
terms $(n_1 \neq n_2)$  the factor $(1/2!)$ is to offset
double counting as in the fermion case. For the diagonal terms,
that vanish in the fermion case, there is no 
double counting but the factor $(1/2!)$ is needed for 
normalization.

By shifting $S_B(T)$ to the left by use of the scattering
formula we see that the series for $Z_B(T)$ is the same
as for $Z_F(T)$ except for the minus signs. Hence
\begin{equation}
Z_B(T) = \exp \left( + \Omega(T) \right)
\end{equation}
where $\Omega(T)$ is defined by the diagram series
in fig 12(c). 

Eq (E10) and E(13) together lead to (E1), the result we
sought to prove here.


\newpage

\epsfxsize=3.0in \epsfbox{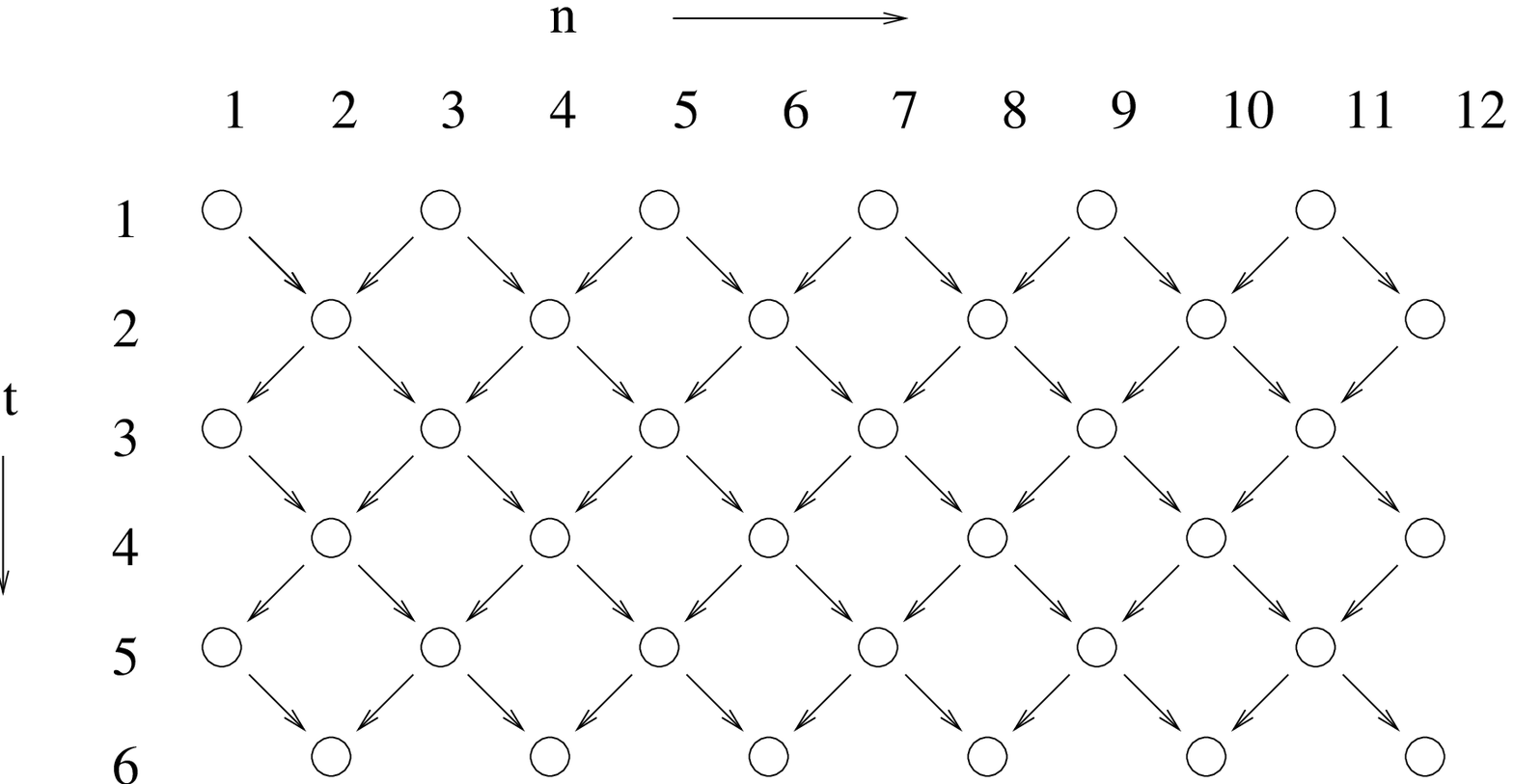}
\figure{Fig 1. The $q$-model of stress propagation through a 
bead pack in 1+1 dimensions. The beads are assumed to sit on 
a regular lattice. Each bead is supported by its two nearest neighbours
in the layer below.}

\vspace{20mm}

\epsfxsize=3.0in \epsfbox{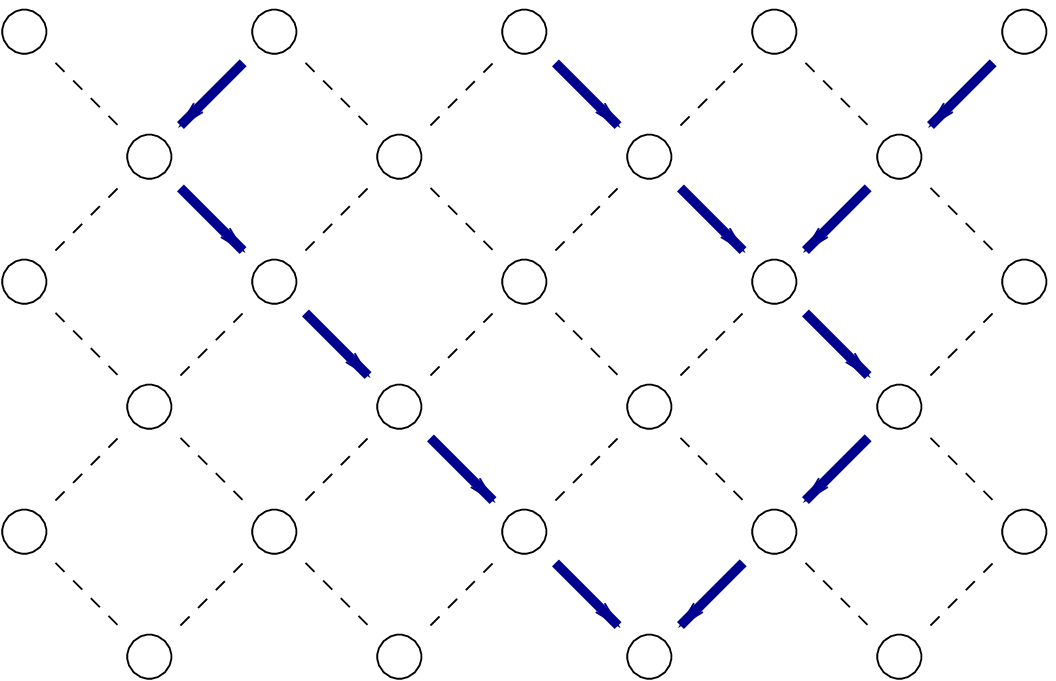}
\figure{Fig 2. {\em Scheidegger's model:} For the
singular $q$-model the load zig-zags down lines that
merge but do not split. In Scheidegger's model of river basins
these lines are interpreted as tributaries merging to form a river.}

\newpage

\epsfxsize=3.0in \epsfbox{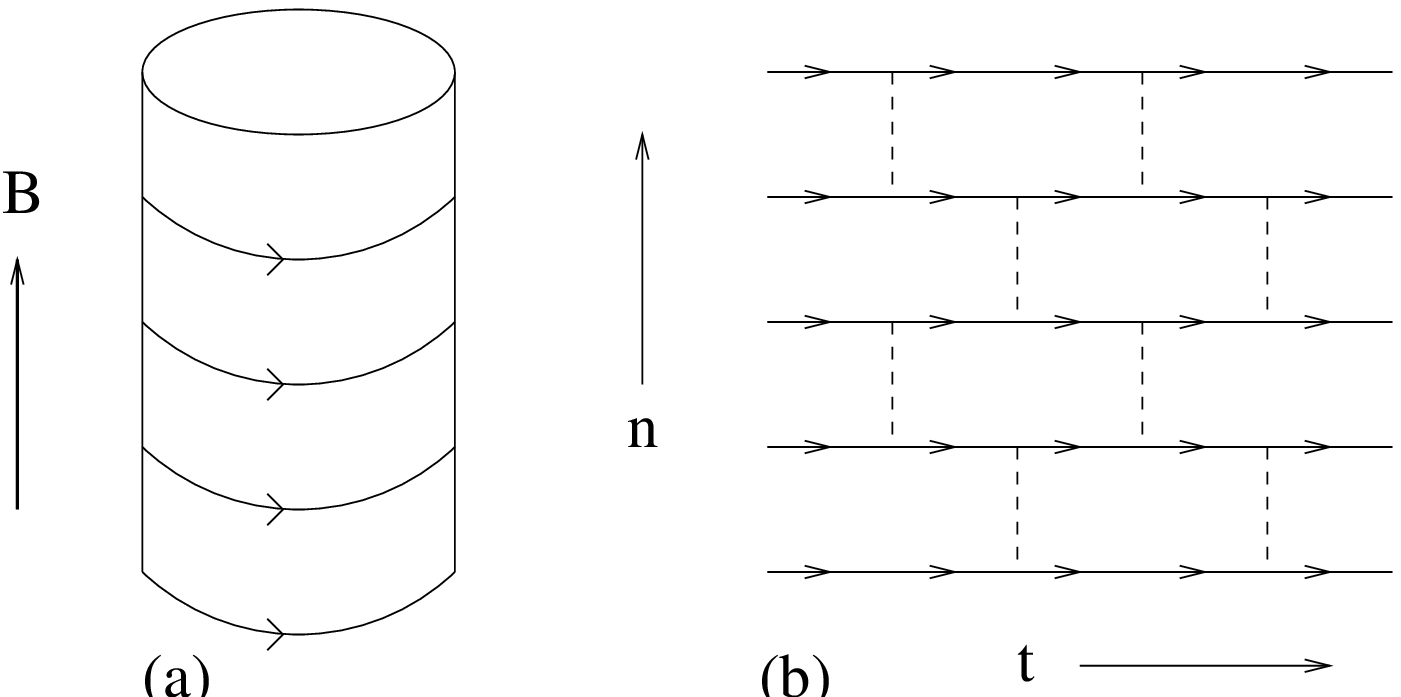}
\figure{Fig 3. {\em A quantum Hall multilayer:} Layers of two dimensional
electron gases are stacked vertically and a strong perpendicular magnetic
field is applied. The important electronic states are at the edge
of each layer. These chiral edge states propagate in the direction
shown in (a). A quantum network model for the surface of the 
multilayer is shown in (b).}

\epsfxsize=3.0in \epsfbox{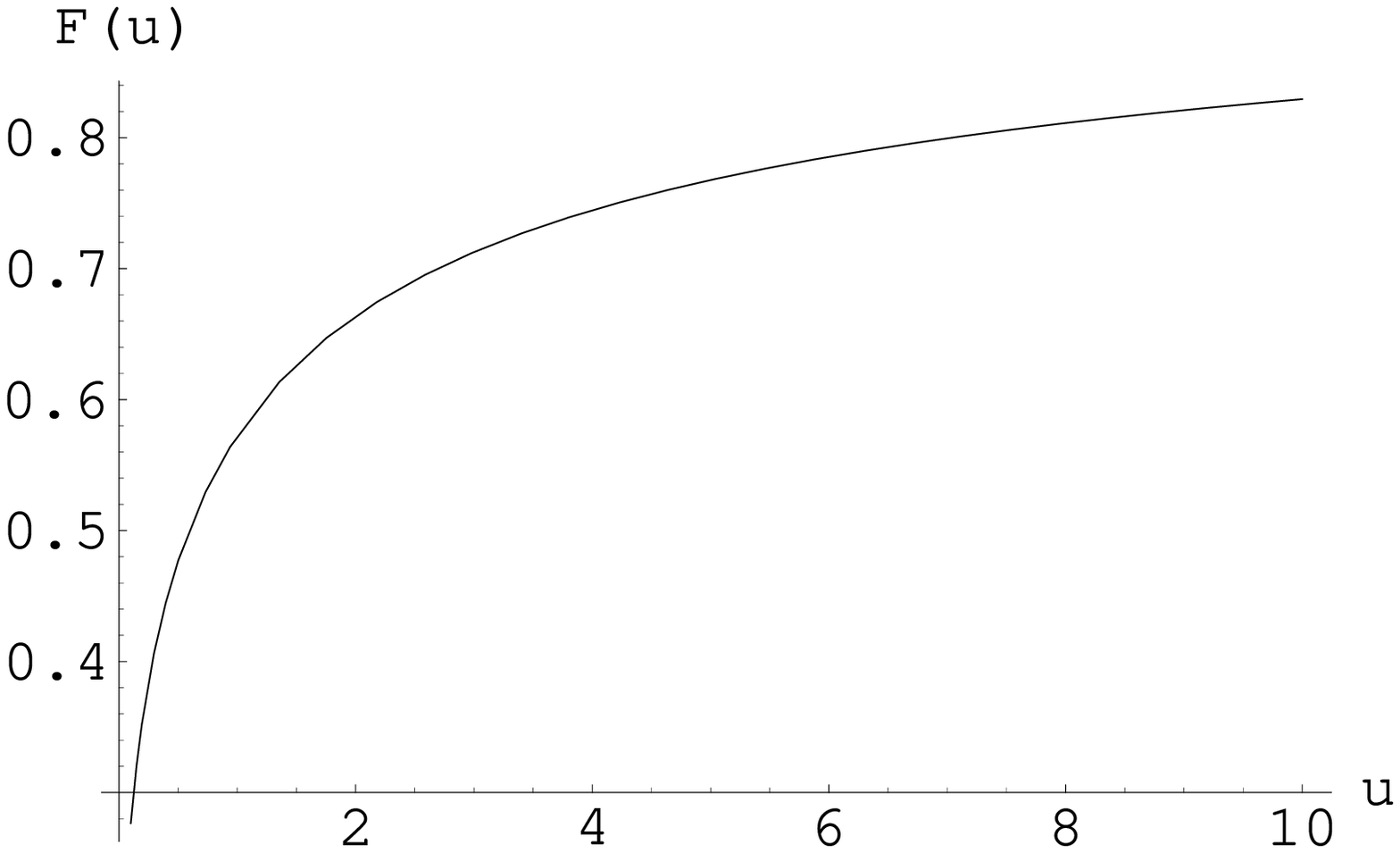}
\vspace{-20mm}
\figure{Fig 4. The scaling function ${\cal F}(u)$ describes the
growth of load fluctuations with depth for the 1+1 dimensional
$q$-model close to the critical point [see eqs (7) and (50)]. Injection
is neglected.}

\newpage

\epsfxsize=3.0in \epsfbox{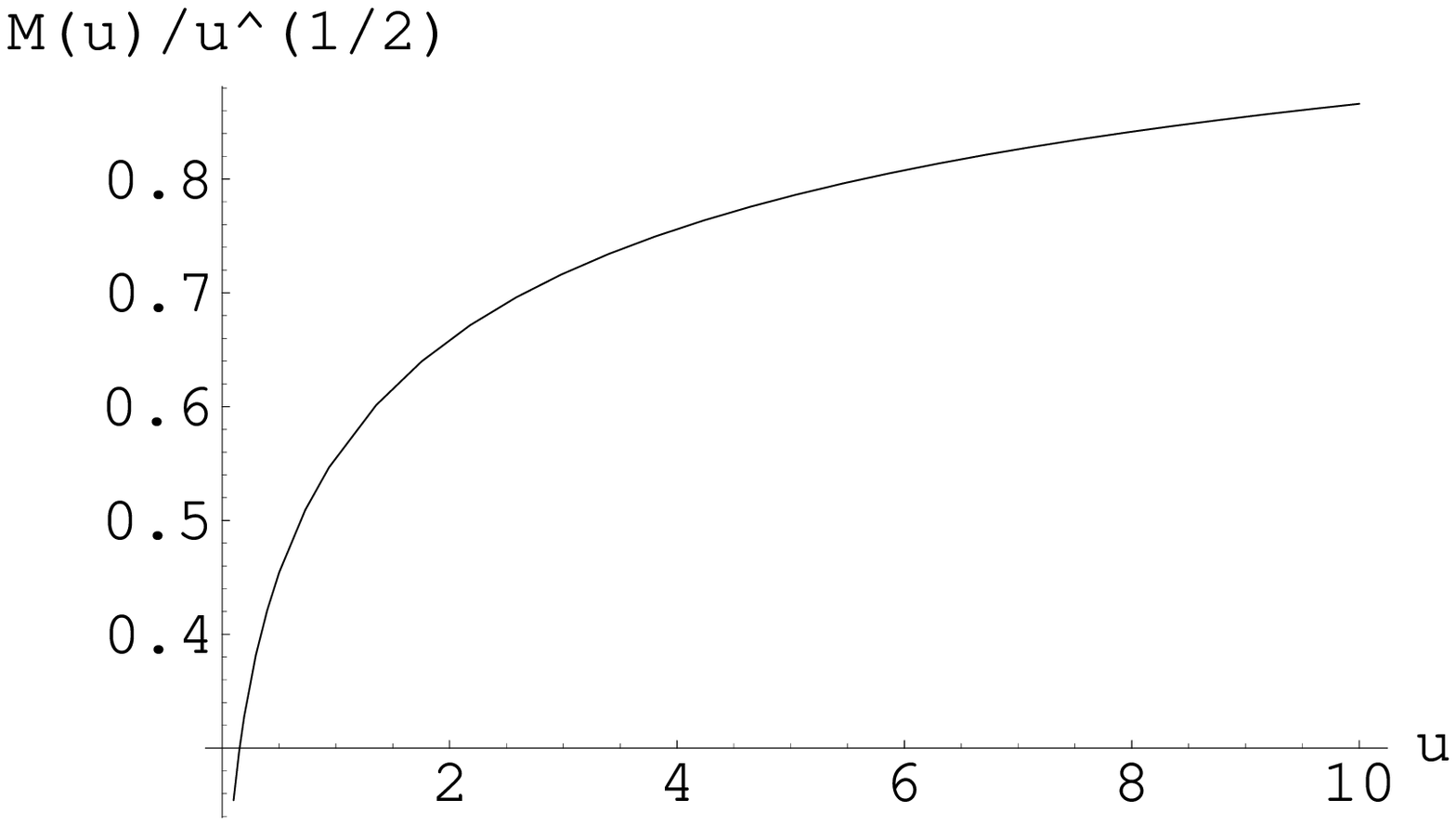}
\vspace{-20mm}
\figure{Fig 5. Growth of load fluctuations with depth for
the 1+1 dimensional $q$-model close to the critical point. The
scaling function ${\cal M}(u)$ gives the contribution due to 
fluctuations in the weight of beads [eqs (79) and (102)]. Here
${\cal M}(u)/\sqrt{u}$ is plotted as a function of $u$.}

\epsfxsize=3.0in \epsfbox{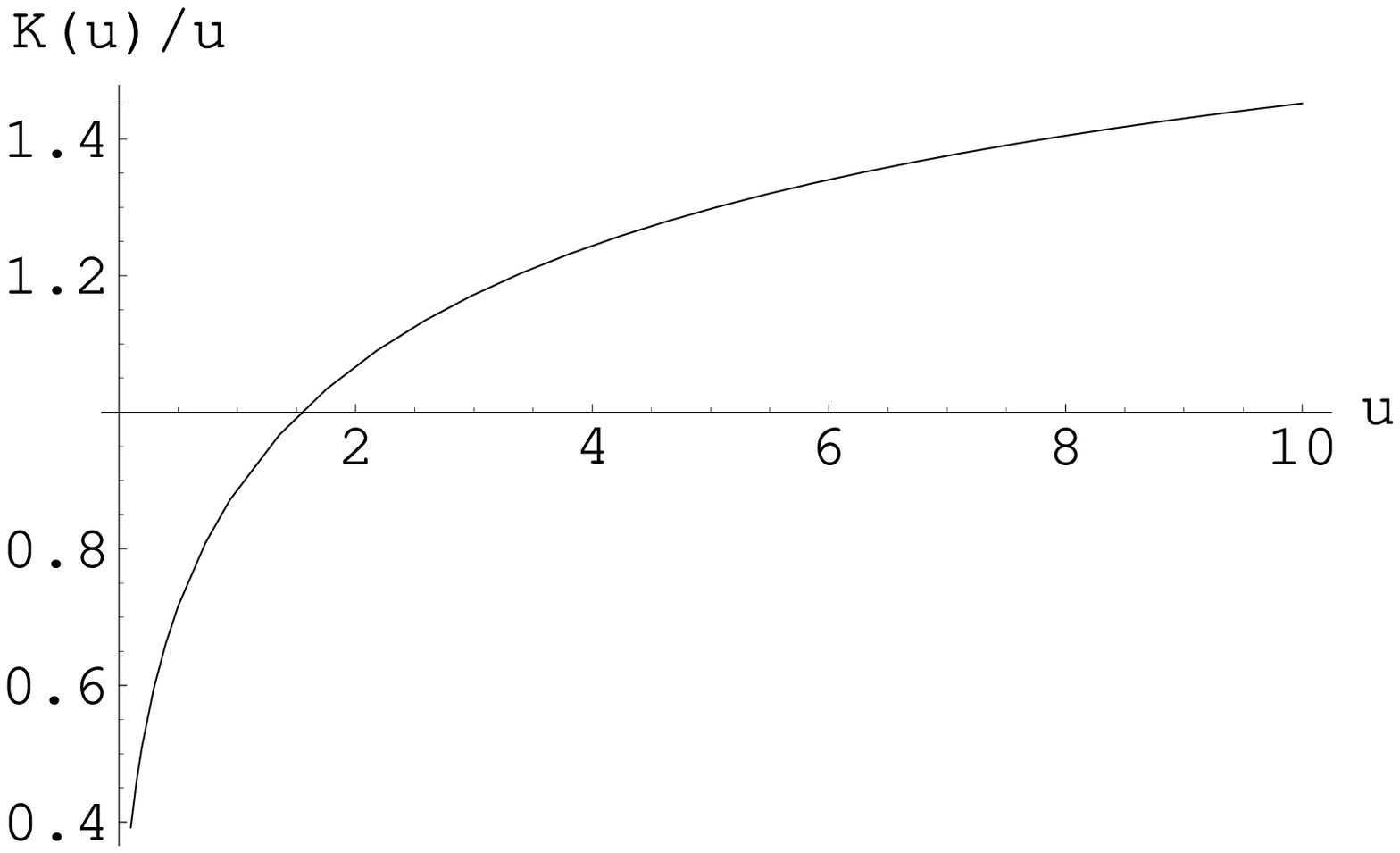}
\vspace{-20mm}
\figure{Fig 6. Growth of load fluctuations with depth for
the 1+1 dimensional $q$-model close to the critical point.
The scaling function ${\cal K}(u)$ gives a contribution proportional
to the average weight of beads [eqs (79) and (102)]. Here
${\cal K}(u)/u$ is plotted as a function of $u$.}

\newpage

\epsfxsize=3.0in \epsfbox{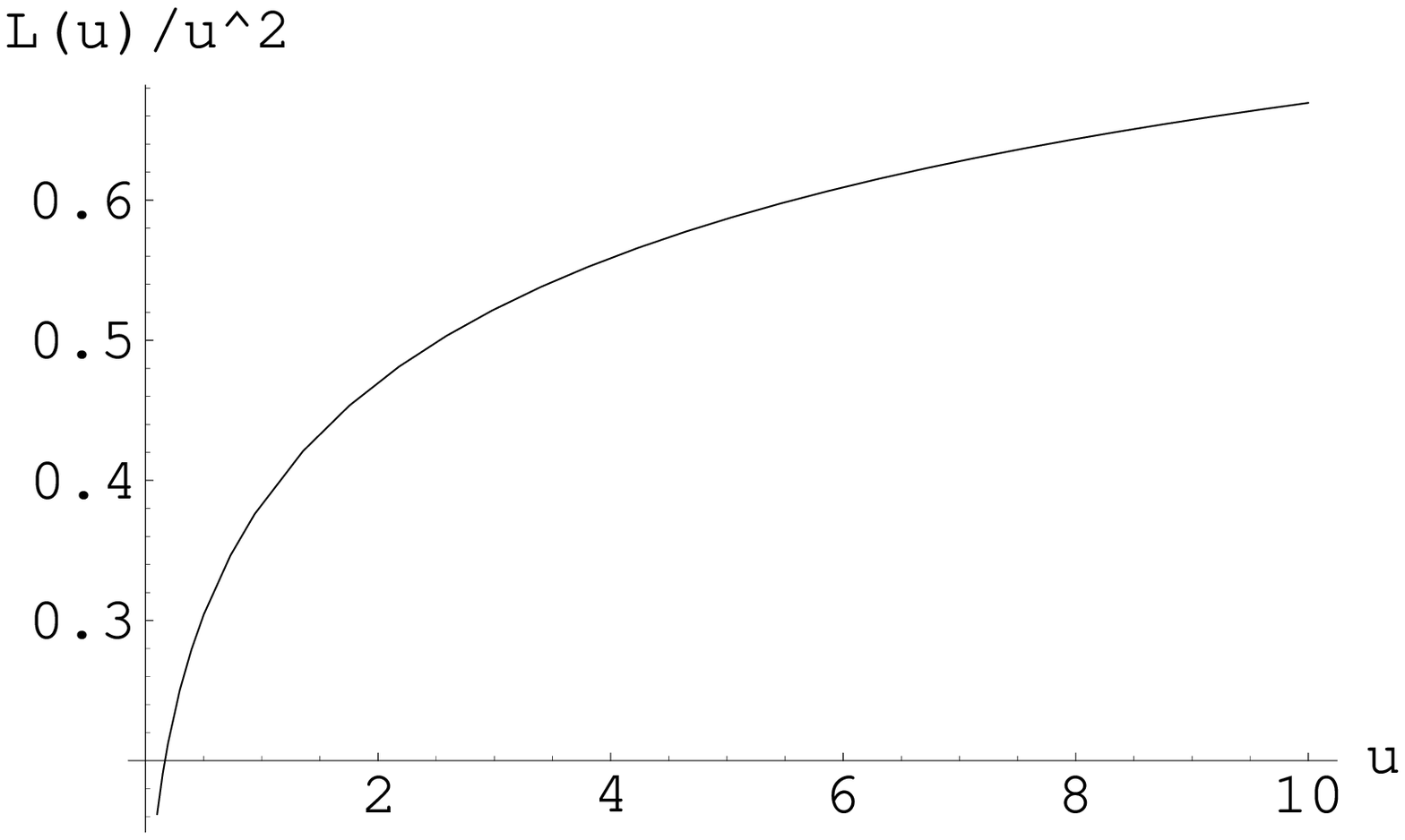}
\vspace{-20mm}
\figure{Fig 7. Growth of load fluctuations with depth for
the 1+1 dimensional $q$-model close to the critical point.
The scaling function ${\cal L}(u)$ gives a contribution proportional
to the square of the average weight of beads [eqs (79) and (102)]. Here
${\cal L}(u)/u^2$ is plotted as a function of $u$.}

\vspace{20mm}

\epsfxsize=3.0in \epsfbox{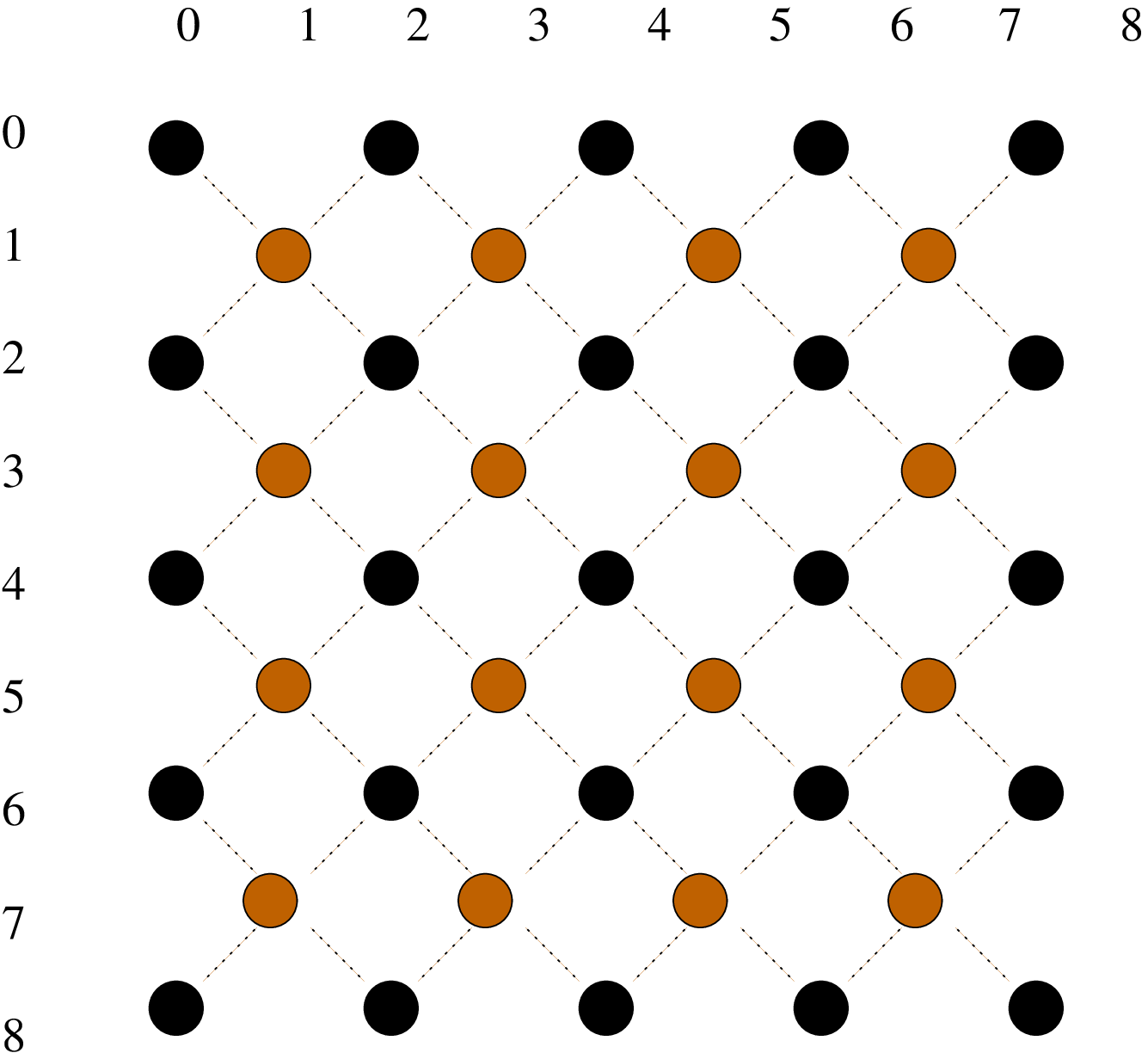}
\figure{Fig 8. Horizontal slice through the 2+1 dimensional
$q$-model. Beads occupy the even (black) or odd (grey) sublattice in 
alternate layers. Each bead is supported by its four nearest neighbours
in the layer below.}

\newpage

\epsfxsize=3.0in \epsfbox{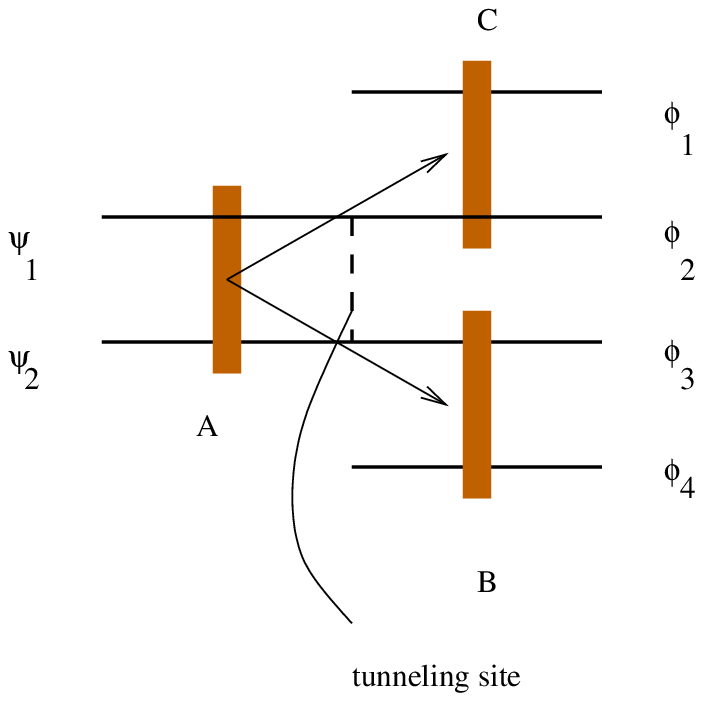}
\figure{Fig 9. Elementary vertex of the Saul, Kardar and Read
model.}

\vspace{20mm}

\epsfxsize=3.0in \epsfbox{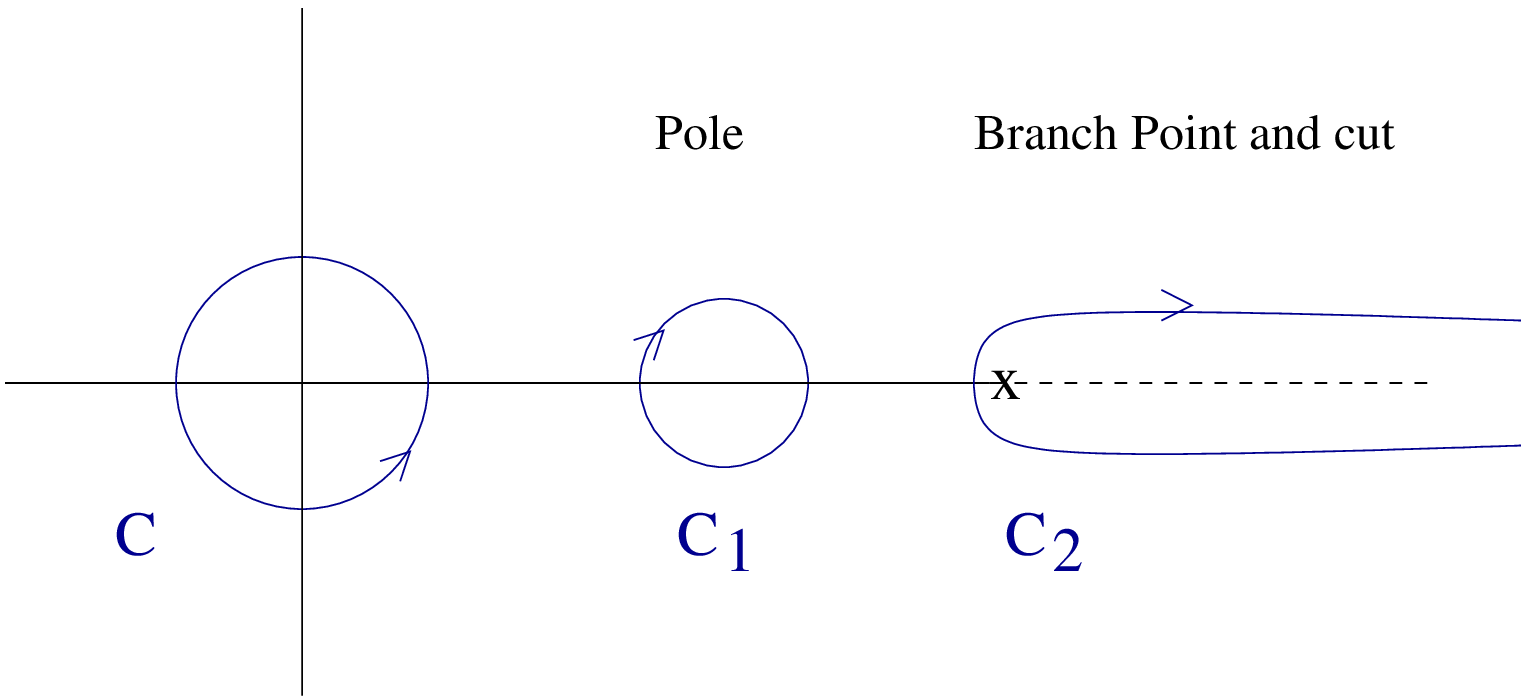}
\figure{Fig 10. Contours for inverting the $z$ transform.}

\newpage

\epsfxsize=3.0in \epsfbox{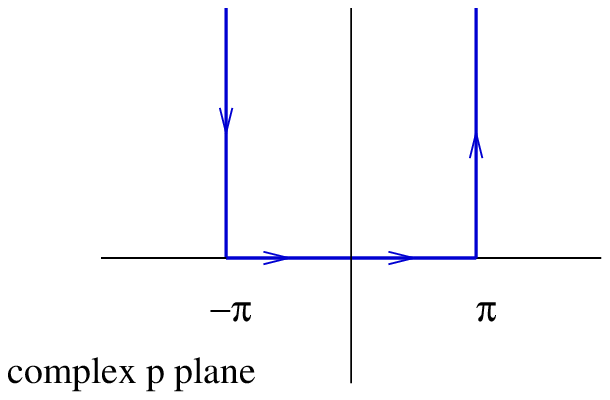}
\figure{Fig 11. Contour for evaluation of two dimensional Green's 
function.}

\epsfxsize=3.0in \epsfbox{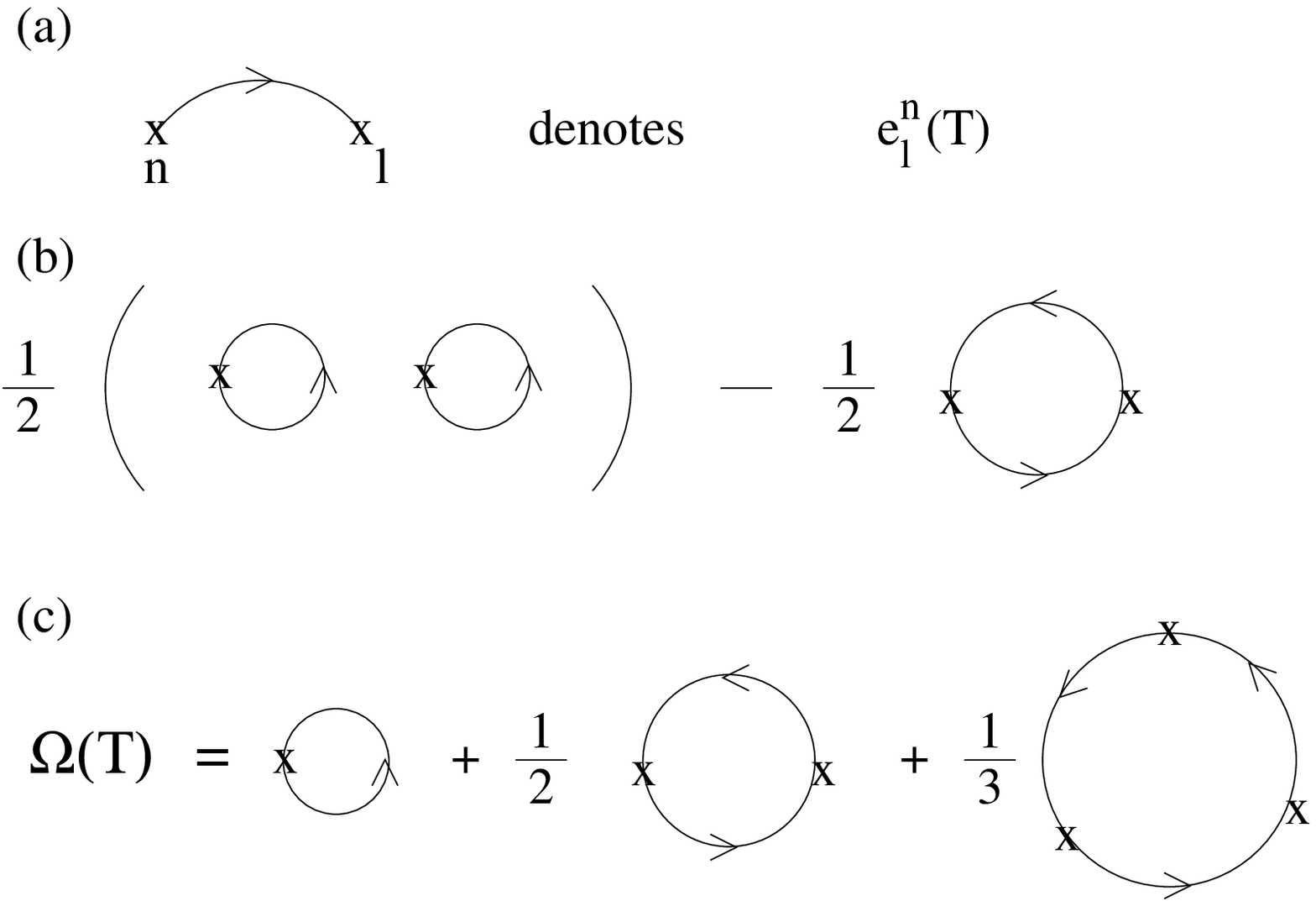}
\figure{Fig 12. Feynman diagrams for the partition function.
(a) Diagram representation of the propagator $e^{n}_{l}(T)$.
(b) Second-order diagrams for the partition function.
(c) Lowest order diagrams in the infinite series for the 
free energy.}

\end{multicols}

\end{document}